%% file: nvc.tex
\documentclass[journal]{IEEEtran}

\usepackage{cite}
\usepackage[pagebackref=true,breaklinks=true,colorlinks,bookmarks=false]{hyperref}       
\usepackage{url}            
\usepackage{float}
\usepackage{graphicx}
\usepackage{subfig}
\usepackage{amsmath,mathrsfs}
\usepackage{amsfonts}
\usepackage{array}
\usepackage{multirow}
\usepackage{flushend}

\begin{document}
%
\title{Neural Video Coding using Multiscale Motion Compensation  and Spatiotemporal Context Model}
%
%
\author{Haojie Liu\textsuperscript{\rm 1}, Ming Lu\textsuperscript{\rm 1}, Zhan Ma\textsuperscript{\rm 1}, Fan Wang\textsuperscript{\rm 2}, Zhihuang Xie\textsuperscript{\rm 2}, Xun Cao\textsuperscript{\rm 1}, and Yao Wang\textsuperscript{\rm 3} \\ \textsuperscript{\rm 1} Nanjing University, \textsuperscript{\rm 2} OPPO.Inc, \textsuperscript{\rm 3} New York University}


%


\maketitle

\begin{abstract}

 Over the past two decades, traditional block-based video coding has made remarkable progress and spawned a series of well-known standards such as MPEG-4, H.264/AVC and H.265/HEVC. On the other hand, deep neural networks (DNNs) have shown their powerful capacity for visual content understanding, feature extraction and compact representation. Some previous works have explored the learnt video coding algorithms in an end-to-end manner, which show the great potential compared with traditional methods. In this paper, we propose an end-to-end deep neural video coding framework (NVC), which uses variational autoencoders (VAEs) with joint spatial and temporal prior aggregation (PA) to exploit the correlations in intra-frame pixels, inter-frame motions and inter-frame compensation residuals, respectively. Novel features of NVC include: 1) To estimate and compensate motion over a large range of magnitudes, we propose an unsupervised multiscale motion compensation network (MS-MCN) together with a pyramid decoder in the VAE for coding motion features that generates multiscale flow fields, 2) we design a novel adaptive spatiotemporal context model for efficient  entropy coding for motion information, 3) we adopt nonlocal attention modules (NLAM) at the bottlenecks of the VAEs for implicit adaptive feature extraction and activation, leveraging its high transformation capacity and unequal weighting with joint global and local information, and 4) we introduce multi-module optimization and a multi-frame training strategy to minimize the temporal error propagation among P-frames. NVC is evaluated for the low-delay causal settings and compared with H.265/HEVC, H.264/AVC and the other learnt video compression methods following the common test conditions, demonstrating consistent gains across all popular test sequences for both PSNR and MS-SSIM distortion metrics.

\end{abstract}


\ifCLASSOPTIONpeerreview
\begin{center} \bfseries EDICS Category: 3-BBND \end{center}
\fi
%
\IEEEpeerreviewmaketitle

\input{intro.tex}
\input{related.tex}

\input{method.tex}

\input{exp.tex}

\input{conc.tex}

\input{ack.tex}
\bibliographystyle{IEEEtran}
\bibliography{nvc}


\end{document}

%% file: intro.tex
{\section{Introduction}\label{sec:introduction}}
\IEEEPARstart{C}{ompressed} video, a dominant media representation across the entire Internet, occupies more than 70\% total traffic volume nowadays for entertainment (e.g., YouTube), productivity (e.g., tele-education), security (e.g., surveillance) etc. 
It still keeps growing explosively. Thus, in pursuit of efficient storage and network transmission, and pristine quality of experience (QoE) with higher resolution content (e.g., 2K, 4K and even 8K video with frame rate at 30 Hz, 60 Hz or even more), a better compression approach is greatly and continuously desired. In principle, the key problem in video coding is how to efficiently exploit visual signal redundancy using prior information, spatially (e.g., intra prediction, transform), temporally (e.g., inter prediction), and statistically (e.g., entropy context adaptation) for more compact representations with less bit rate consumption  at the same reconstruction quality. This is well formulated as the minimization of Lagrangian cost $J$ of rate-distortion optimization (RDO) that is widely adopted in existing video coders, e.g.,
\begin{align}
  \min J = R + \lambda\cdot D, \label{eq:rdo}
\end{align} with $R$ and $D$ represent the compressed bit rates and reconstructed distortion respectively.

{\bf Motivation.} Over the past three decades,  video compression technologies have been evolving and adapting constantly with coding efficiency improved by several folds, mostly driven under the efforts from the experts in ISO/IEC Moving Picture Experts Group (MPEG), ITU-T Video Coding Experts Group (VCEG) and their joint task forces. It leads to  several popular video coding standards, including the H.264/Advanced Video Coding (H.264/AVC)~\cite{wiegand2003overview}, High-Efficiency Video Coding (HEVC)~\cite{sullivan2012overview} and emerging versatile video coding (VVC)~\cite{VVC}. These standards share the similar (recursive) block-based hybrid prediction/transform framework where individual coding tools, such as the intra/inter prediction, integer transforms,  context-adaptive entropy coding, etc, are intensively handcrafted to optimize the overall efficiency. Among them, {\it pixel-domain  predictive coding} is one of the most important factors, contributing to the major performance gains~\cite{ohm2012comparison}. For example, pixel-domain intra prediction was officially adopted into the H.264/AVC and later extended with the support of recursive block-sizes and abundant predictive directions for efficiently exploiting spatial structures; recursive and even non-squared blocks are extensively used in inter prediction to remove temporal coherency. Basically, conventional video coding methods leverage the spatiotemporal pixel neighbors (as well as their linear combinations) for predictive signal construction, resulting in corresponding  residuals for subsequent  transform, quantization, and entropy coding for more compact representation. Optimal coding mode with appropriate block size and orientation (e.g., intra direction, inter motion vectors) is selected via computational RDO process, utilizing $\ell1$-norm (e.g., mean absolute error - MAE) or $\ell2$-norm (e.g., mean squared error - MSE) as the distortion 
metric.  

Though recursive block-based pixel prediction shows its great success, it is mainly due to the hardware advancements in past decades, by which we can exhaustively search for the best prediction. It, however, is more and more challenging to simply trade computational resources for efficiency improvement because Moore's Law does not hold any more~\cite{dean20201}. It therefore calls for innovative methodologies and architectures of video coding to further improve the coding efficiency, in response to the ever increasing users' requirement for video resolution and quality. Such pixel prediction strategy, either intra or inter, mostly relies on the physical coherence of video signal and applies the mathematical tools (e.g., linear weighting, orthogonal transform, Lagrangian optimization) for signal energy compaction.  

\begin{figure}[t]
     \centering
     \includegraphics[scale=0.37]{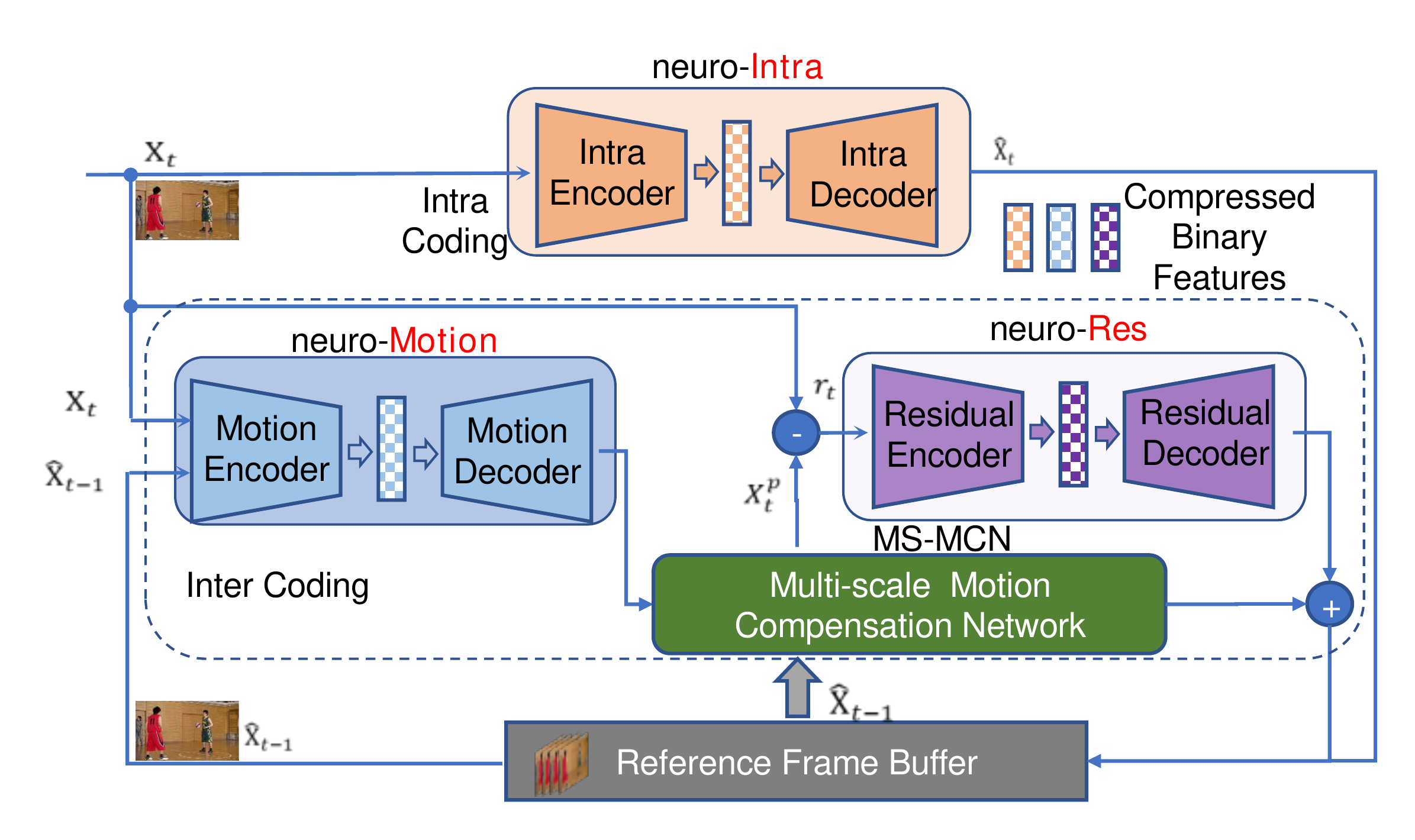}
     \caption{{\bf Neural Video Coding (NVC).} The modules neuro-Intra, neuro-Res, and neuro-Motion follow the general model architecture in Fig.~\ref{fig:bio_video_coder} for efficient representations of intra pixels, displaced inter residuals, and inter motions. The neuro-Motion uses a pyramid decoder for the main decoder as discussed in Sec.~\ref{ssec:ms_mrc}.}
     \label{fig:nvc_architecture}
\end{figure}

{\bf Our Approach.} We propose an end-to-end neural video coding framework (NVC), which codes intra-frame pixels (called neuro-Intra), inter-frame motion (called neuro-Motion), and inter-frame residual (called neuro-Res) using separate variational autoencoders (VAE), as shown in Fig.~\ref{fig:nvc_architecture}. A multiscale motion compensation network (MS-MCN) works together with neuro-Motion to generate multiscale optical flows and perform multiscale motion-compensated prediction of the current frame from the previous frame. The sparse image differences between past and present frame, e.g., residuals, are then encoded to obtain the final reconstruction; All three VAEs, e.g., neuro-Intra, neuro-Motion, neuro-Res for compressing intra-pixel, inter-motion and inter-residual, are engineered together with MS-MCN in an end-to-end learning manner. Note that   neuro-Intra takes a native image frame as input, neuro-Motion uses the current and past reconstructed frame, generating multiscale compressed flows (MCFs), MS-MCN uses these generated  MCFs for motion compensation to obtain the inter-predicted frame  and neuro-Res encodes the difference between the current frame and its prediction for the final reconstruction. Additionally, joint spatiotemporal and hyper priors are aggregated for efficient and adaptive context modeling of latent features to improve entropy coding efficiency for the motion field.

We have evaluated the efficiency of the proposed NVC for the low-delay causal settings against well-known HEVC, H.264/AVC and  other learnt video compression methods following the common test conditions. The NVC demonstrated the leading performance with consistent gains across all popular test sequences for both PSNR (Peak signal-to-noise ratio) and MS-SSIM (multiscale structural similarity)~\cite{wang2003multiscale} distortion metrics. Using the H.264/AVC as a common anchor, our NVC presents 35\% BD-Rate (Bjontegaard Delta Rate)~\cite{bjontegaard2001calculation} gains, while HEVC and DVC (Deep Video Coding)~\cite{Lu_2019_CVPR} offer 30\% and 22\% gains, respectively, when the distortion is measured in terms of PSNR. Gains are even larger, if the distortion metric is replaced by the MS-SSIM. 
In this case, NVC can achieve 50\% improvement, while both HEVC and DVC are around 25\%. We further compare our NVC with DVC\_Pro~\cite{lu2020end} (an upgraded version of DVC) on HEVC test sequences: our NVC reveals 32.20\% and 50.07\% BD-Rate reduction measured by PSNR and MS-SSIM distortion respectively, while DVC\_Pro gives 34.57\% and 45.88\%.

Ablation studies have also been conducted to examine the gains due to different modules of NVC. We have shown that temporal priors and multi-frame training could greatly improve the efficiency and learning stability. Our MS-MCN is able to remove motion compensation noise by  multiscale compensation, even better than using a cascaded trained denoising network.

\begin{table}[t]
    \centering
    \caption{Abbreviations and Notations}
    \begin{tabular}{c|c}
    \hline
        abber. & description \\
        \hline
         NLAM & NonLocal Attention Module\\
         LAM & Local Attention Module\\
         MS-MCN & Multi-scale  Motion Compensation Network\\
         SS-MCN & Single-scale Motion Compensation Network\\
         PA & Prior Aggregation\\
         MCF & Multiscale Compressed Flow\\
         neuro-Intra &  Neural intra coding\\
         neuro-Motion & Neural motion coding\\
         neuro-Res & Neural residual coding\\
         \hline
         MSE & Mean Squared Error\\
         MAE & Mean Absolute Error\\
         PSNR & Peak signal-to-noise ratio\\
         MS-SSIM & Multiscale Structural Similarity\\
         \hline
         
    \end{tabular}
    \label{tab:annotation}
\end{table}

\begin{figure*}[t]
\centering
\includegraphics[scale=0.42]{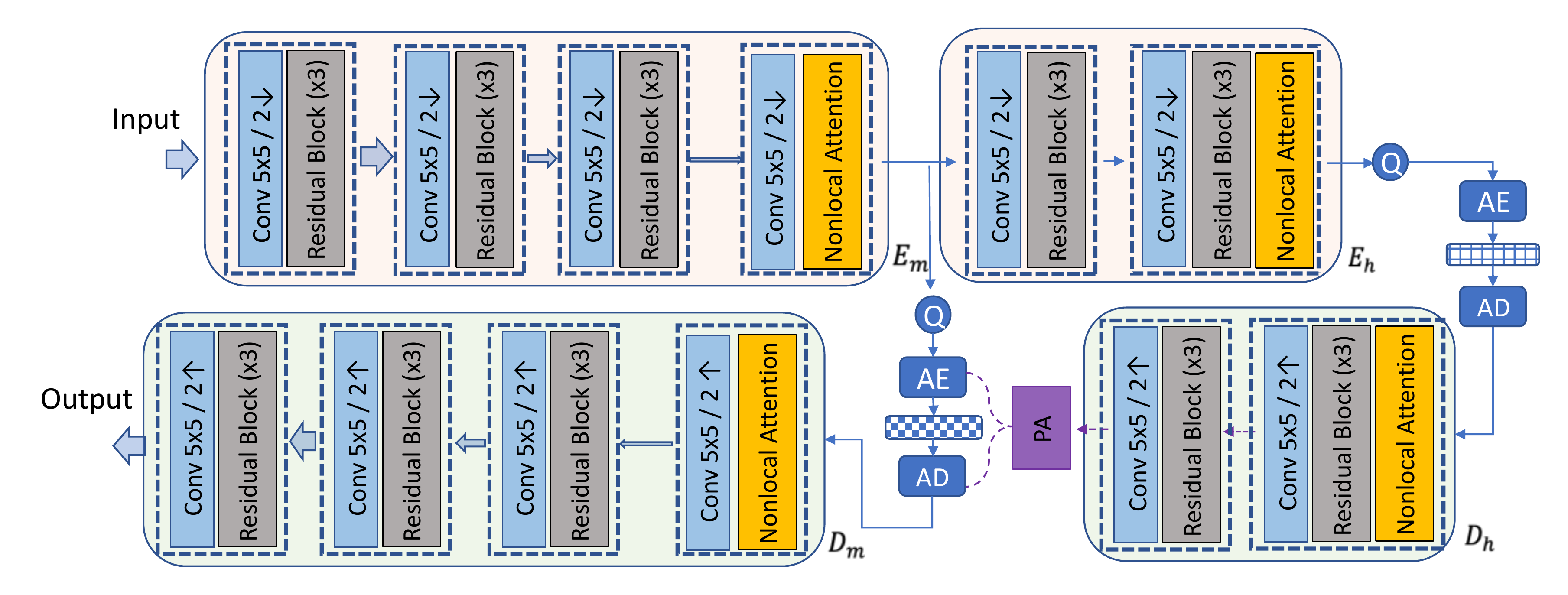}
\caption{{\bf The general structure of the variational autoencoder based compression engine used for neuro-Intra, neuro-Motion and neuro-Res in Fig.~\ref{fig:nvc_architecture}.} The main encoder ${\bf E}_m$ includes for major convolutional layers (each including a convolutional layer followed by down-sampling and three residual blocks, and the last layer includes a non-local attention module). The hyper encoder ${\bf E}_h$ includes two major convolutional layers. The main decoder ${\bf D}_m$ and hyper decoder ${\bf D}_h$ reverse the processing of ${\bf E}_m$ and ${\bf E}_h$, respectively. Prior aggregation (PA) engine collects the information from hyper prior, autoregressive spatial neighbors, as well as temporal priors (if applicable) for efficient modeling of the probability distribution of latent features generated by the main encoder. Nonlocal attention is adopted at the bottlenecks of both main and hyper encoders to enable saliency based bit allocation, and rectified linear unit (ReLU) is embedded with convolutions for enabling the nonlinearity. ``Q'' is for quantization, AE and AD are for arithmetic encoding and decoding, respectively. 2$\downarrow$ and 2$\uparrow$ are downsampling and upsampling at a factor of 2 for both horizontal and vertical dimensions.}
\label{fig:bio_video_coder}
\end{figure*}

{\bf Novelty.} The main contributions are highlighted below:
\begin{itemize}
\item We propose an end-to-end deep neural video coding framework (NVC), leveraging learnt feature domain representations  for intra-pixel, inter-motion and inter-residual, respectively for compression;

\item neuro-Motion and multiscale motion compensation network (MS-MCN) are employed together to capture coarse-to-fine motion displacements and obtain the prediction by warping the features of the reference frame at multiple scales;
\item We propose a novel spatiotemporal context modeling approach for the entropy coding of the motion features, where the temporal context is obtained through modeling the temporal evolution of the motion features using a ConvLSTM in the temporal updating module (TUM). The temporal context is combined with the autoregressive spatial context and hyperprior features in a spatiotemporal hyper aggregation module (STHAM);
\item Nonlocal attention is attached at bottleneck layers of the VAE  modules for adaptive bits allocation based on joint global and local feature extraction implicitly.
\end{itemize}
 
 This work is based on our preliminary work~\cite{liu2019learned} but with significant extensions and discussions including pyramid flow decoder in neuro-Motion, replacing single scale motion compensation with multiscale motion compensation, spatiotemporal prior aggregation engine for context modeling and progressive training with multiple frames.

 The  rest  of  this  paper  is  structured  as  follows: Sec.~\ref{sec:related_work} will briefly review relevant studies about leaned image and video coding. Our neural video coding framework (NVC) is given in Sec.~\ref{sec:nvc} with detailed discussions about neuro-Intra, neuro-Motion, MS-MCN and neuro-Res; Sec.~\ref{sec:exp} will present experiments and ablation studys; and  concluding  remarks  are  drawn  in  Sec.~\ref{sec:conc}.


%% file: related.tex
\section{Related Work} \label{sec:related_work}
Built on advancements of deep neural networks (DNNs), we have seen the explosive growth of DNN-based image/video compression approaches. Some explorations have attempted to replace modular components in traditional image/video coding framework such as filtering, intra prediction, etc; and some others have fully relied on powerful learning tools to perform the end-to-end optimization. Given that our neural video coding framework (NVC) belongs to the end-to-end learning category, we will emphasize the reviews in this avenue. For the modular optimization using DNNs, two great review articles can be found in~\cite{liu2020deep,ma2019image}.

\subsection{Learnt Image Compression}


DNN-based image compressions usually utilize auto-encoder or VAE architectures, consisting of nonlinear transforms, attention or importance map, differentiable quantization, context model, and embedded loss functions, for end-to-end learning. 

{\bf Recurrent Autoencoder.} Toderic et al.~\cite{toderici2015variable} first proposed to utilize fully-connected recurrent auto-encoders for variable-rate thumbnail image compression. A serial improvements were then extended, including the full-resolution image support, learned entropy coding, unequal bits allocation, etc~\cite{toderici2017full,johnston2018improved} by the introductions of  ConvLSTM or ConvGRU, for better coding efficiency. Variable bit rate is intrinsically enabled by such recurrent structure. It, however, suffers from the higher computational complexity at higher bit rates, because  more recurrent processing are desired.

{\bf Convolutional Autoencoder.} Alternatively, convolutional auto-encoders~\cite{balle2016end,balle2018variational,mentzer2018conditional,minnen2018joint}, are extensively studied in past years where different bit rates are adapted by setting a variety of $\lambda$ in learning to optimize~\eqref{eq:rdo}.  Note that different network models may be required for 
individual bit rates, making it difficult for hardware implementation (e.g., model adaptation for various bit rates). Recently, conditional convolution~\cite{choi2019variable} and scaling factor~\cite{chen2019neural} were proposed to enable variable-rate compression using a single or a limited number of network parameters without noticeable coding efficiency loss. It makes the convolutional autoencoders more attractive for practical applications.

{\bf Attention/Importance Map.} Li et al.~\cite{li2017learning} utilized a separate three-layer CNN to generate importance map for spatial-complexity-based adaptive bits allocation, leading to the noticeable subject quality improvement with well-preserved edges and textures. Instead, Mentzer et al.~\cite{mentzer2018conditional} further selected one channel from the bottleneck layer to unequally weigh features at different spatial locations for simplification. Such importance map embedding was lightweight and easy for training and end-to-end optimization.
This approach was later improved with nonlocal attention mechanism to efficiently and implicitly capture both global and local important information for better compression~\cite{chen2019neural}.

{\bf Nonlinear Transform.} To generate more compact feature representation, Balle et al.~\cite{balle2016end} suggested to replace the traditional nonlinear activation, e.g., ReLU, with the generalized divisive normalization (GDN) that was theoretically proven to be more consistent with image natural statistics for visual perception. A succeeding investigation by the same authors was given in~\cite{balle2018efficient}, reporting that GDN outperformed other nonlinear rectifiers, such as ReLU, leakyReLU and tanh, in compression tasks. Several follow-up studies~\cite{lee2018context,klopp2018learning} directly applied GDN in their networks for compression exploration.


{\bf Adaptive Contexts.} Probabilistic model plays a vital role in data compression. Assuming the Gaussian distribution for feature elements, Balle et al.~\cite{balle2018variational}  utilized hyper priors to estimate the parameters of Gaussian Scale Model (GSM) for latent features. Later Hu et al.~\cite{hu2020coarse} used hierarchical hyper priors (coarse-to-fine) for improving the entropy models in multiscale representations.  Minnen et al.~\cite{minnen2018joint} improved the context modeling using joint autoregressive spatial neighbors and hyper priors based on Gaussian Mixture Model (GMM).  Autoregressive spatial priors were usually extracted by PixelCNNs or PixelRNNs~\cite{oord2016pixel}, which have been widely adopted for natural image density modeling. Reed et al.~\cite{reed2017parallel} further introduced multiscale PixelCNNs, yielding competitive density estimation and great speedup (e.g., from $O(N)$ to $O(\log N)$). It was later extended from 2D architectures to 3D PixelCNNs~\cite{mentzer2018conditional}. Channel-wise weights sharing-based 3D implementations can greatly reduce network parameters with higher parallelization. Then Chen et al.~\cite{chen2019neural} discussed parallel pipelines of 3D PixelCNNs for practical decoding. Previous methods accumulated all the accessible priors to estimate the probability based on a single Gaussian distribution for each element. Recent explorations have shown that weighted GMMs can further improvre the coding efficiency as reported by~\cite{cheng2020learned,lee2019end}.



{\bf Quantization.} Quantization is a non-differentiable operation, basically converting continuous variables into discrete variables with a limited alphabet. This process has to be replaced by a differentiable operation when used in end-to-end learning framework for back propagation. A number of methods, such as uniform noise adding~\cite{balle2016end}, stochastic rounding~\cite{toderici2015variable}, soft-to-hard vector quantization~\cite{mentzer2018conditional} and universal quantization~\cite{choi2019variable}, were developed to approximate a continuous distribution for differentiation. 


{\bf Loss Functions.} Pixel-error, such as MSE, or MAE, was one of the most popular loss functions used. In the meantime, SSIM or MS-SSIM was also adopted because of its better consistency with visual perception. Simulations had revealed that SSIM-based loss can improve the perception quality, especially at low bit rates.
Towards the perception-optimized encoding, perceptual losses that were measured by adversarial loss~\cite{rippel2017real,huang2019extreme,agustsson2019generative} and VGG loss~\cite{liu2018deep} were embedded in learning to produce visually appealing results.

\subsection{Learnt Video Compression}

Learnt video compression is extended from the learnt image compression by further exploiting the temporal redundancy in a trainable way for efficient representations of temporal motion and displaced image difference (e.g., predictive residual). In most cases, network models for image compression were directly re-used for temporal displaced residuals, leaving a great deal of efforts devoted for better motion field compression.

 Chen et al.~\cite{chen2017deepcoder} developed the DeepCoder where a simple convolutional autoencoder was applied for both intra and residual coding at fixed 32$\times$32 blocks, and block-based motion estimation in traditional video coding was re-used for temporal compensation. Lu et al.\cite{Lu_2019_CVPR} introduced the optical flow for motion representation in their DVC work, which, together with the intra coding in~\cite{balle2018variational}, demonstrated similar performance compared with the HEVC. However, the coding efficiency suffer from a sharp loss at low bit rates. Liu et al.~\cite{liu2019learned} extended their nonlocal attention optimized image compression (NLAIC) for coding the intra and residual frame, and applied spatiotempoal adaptive context model for more compact motion representation, showing consistent rate-distortion gains across different contents and bits rates. 

Motion can be also implicitly inferred by temporal interpolations. For example, Wu et al.~\cite{wu2018vcii} applied recurrent neural network (RNN)-based frame interpolation. Together with the residual compensation, it offered comparable performance with H.264/AVC. Djelouah et al.~\cite{djelouah2019neural} further imporved the interpolation-based video coding by utilizing the advanced optical flow estimation and feature domain residual coding. Yang et al.~\cite{Yang_2020_CVPR} also use a interpolation method with three hierarchical quality layers and a recurrent enhancement network for compression.
However, temporal interpolation usually led to coding delay that may not be acceptable for low-latency applications.

Another interesting exploration made by Ripple et al. in~\cite{rippel2019learned} was to jointly encode the flow and residual signals using unified quantized features in an unsupervised way. A recurrent state was embedded to aggregate multi-frame information for efficient flow generation and residual coding.

%% file: method.tex
\section{Neural Video Coding} \label{sec:nvc}

\subsection{Overview of NVC}

 Our NVC framework is designed for low-delay applications. As with all modern video encoders, the proposed NVC compresses the first frame in each group of pictures as an intra-frame using a VAE based compression engine (neuro-Intra).  It codes remaining frames using motion compensated prediction. As shown in Fig.~\ref{fig:nvc_architecture}, it uses the VAE compressor (neuro-Motion) to generate the multiscale motion field between the current frame and the reference frame. Then, MS-MCN takes multiscale compressed flows, warps the multiscale features of the reference frame, and combines these warped features to generate the predicted frame.  The prediction residual is then coded using another VAE-based compressor (neuro-Res).
 
 Given a group of pictures (GOP) $\mathbb{X}$ = \{${\bf X_1},{\bf X_2},...,{\bf X_t}$\}, we first encode ${\bf X_1}$ using the neuro-Intra module, leading to the reconstructed frame $\hat{\bf X}_1$. The following frame ${\bf X}_2$ is encoded predictively, using neuro-Motion, MS-MCN, and neuro-Res together, shown in Fig.~\ref{fig:nvc_architecture}.  
Note that MS-MCN takes the multiscale optical flows $\left\{\vec{f}^1_d,\vec{f}^2_d,...,\vec{f}^s_d\right\}$ derived by the pyramid decoder in neuro-Motion, and then generates the predicted frame $\hat{\bf X}^p_2$ by multiscale  motion compensation. Displaced inter-residual $ {\bf r}_2 = {{\bf X}_2} - {\hat{\bf X}^p_2}$ is then compressed in neuro-Res, yielding the reconstruction $\hat{\bf r}_2$. The final reconstruction $\hat{\bf X}_2$ is given by
${\hat{\bf X}_2} = {\hat{\bf X}^p_2} + {\hat{\bf r}_2}$.  Encoding of the next P-frame  follows the same procedure of ${\bf X}_2$ until all frames in the GOP are coded completely. 

Table~\ref{tab:annotation} summarizes relevant abbreviations used throughput this paper.
\subsection{The VAE Architecture using NLAM and Spatiotemporal Priors}
The general architecture of the VAE model is shown in Fig.~\ref{fig:bio_video_coder}, with main encoder-decoder pair for latent feature analysis and synthesis, and hyper encoder-decoder for hyper prior generation. The main encoder ${\bf E}_m$ uses four stacked CNN layers, where each convolutional layer employs stride convolutions to achieve downsampling (e.g., at a factor of 2 in this example) and cascaded convolutions (e.g., three ResNet-based residual blocks~\cite{he2016deep}\footnote{We apply cascaded ResNets for stacked CNNs because of its high-efficiency and reliable performance. Other efficient CNNs architectures can be applied as well.}) for efficient feature extraction. We utilize two-layer hyper encoder  ${\bf E}_h$ to further generate the subsequent hyper priors as side information, which are used for the entropy coding for the latent features.


To capture the spatial locality, we apply convolutional layers with limited receptive field (e.g., 3$\times$3) that are stacked altogether to simulate the layer-wise feature extraction. These same ideas are used in many relevant studies~\cite{balle2018variational,minnen2018joint}. We utilize the simplest ReLU as the nonlinear activation function. Other nonlinear activation functions can be used as well, such as the GDN (Generalized Divisive Normalization)  in~\cite{balle2016end}. 

Attention mechanism is leveraged to intelligently allocate bit resource (e.g., via unequal feature quantization) for image/video compression~\cite{li2018learning,mentzer2018conditional}. It basically enables more accurate reconstruction of salient areas. We adopt the nonlocal attention module (e.g., NLAM) at the bottleneck layers of both the main encoder and hyper encoder, prior to quantization to include both global and local information for more accurate importance selection. This module is motivated by the behaviour of HVS, where we often promptly scan the entire viewing scene to have the complete understanding of the field of vision, and then fixate to the salient regions. 

To enable more accurate conditional probability density  modeling for entropy coding of the latent features, we introduce the {\it Prior Aggregation} (PA) engine which fuses the inputs from the hyper priors, spatial neighbors and temporal context (if applicable)\footnote{Intra and residual coding only use joint spatial and hyper priors without temporal inference.}. The more accurate context modeling requires less resource (e.g., bits) for information representation as suggested in information theory~\cite{cover2012elements}. For the sake of simplicity, we assume the latent features (e.g., motion, image pixel, residual) following the Gaussian distribution as in~\cite{minnen2018joint,hu2020coarse}, and use the PA engine to derive the mean and standard deviation of the distribution for each feature.

\subsection{Neural Intra Coding}
\label{sec:intra_coding}
Our neuro-Intra is a simplified version of the NLAIC that was originally proposed in \cite{chen2019neural}.

{\bf {NLAIC:}} One major distinction of NLAIC from the VAE model using autoregressive spatial context in \cite{minnen2018joint} is the introduction of a nonlocal attention module (NLAM) inspired by~\cite{zhang2019residual}. NLAM is used to capture the global and local importance for saliency aggregation, by which we can assign different importance to various spatial-channel elements implicitly. Such nonlocal attention mechanism is inspired by the visual information processing of spatial perception (e.g., coarse global structure plus fine-grain local details). We have found that with the addition of NLAM, we can achieve similar performance as \cite{minnen2018joint} without using GDN nonlinearity.

In addition, we have applied 3D 5$\times$5$\times$5 masked CNN\footnote{This 5$\times$5$\times$5 convolutional kernel shares the same parameters for all channels, offering great model complexity reduction compared with 2D CNN-based solution in~\cite{minnen2018joint}.} to extract spatial priors, which are fused with hyper priors in PA for entropy context modeling (e.g., bottom part of Fig.~\ref{fig:entropy_model}).
Here, we have assumed the single Gaussian distribution for the context modeling of entropy coding. More details of NLAIC can be found in~\cite{chen2019neural}. Note that temporal priors are not adopted for intra-pixel and inter-residual in this paper.


{\bf {Improvements to NLAIC.}} Original NLAIC applies multiple NLAMs in both main and hyper coders, leading to excessive memory consumption at a large spatial scale. In NVC, NLAMs are only used at the bottleneck layers for both main and hyper encoder-decoder pairs, achieving adaptive bits allocation implicitly.

\begin{figure}[t]
     \centering
     \includegraphics[scale=0.30]{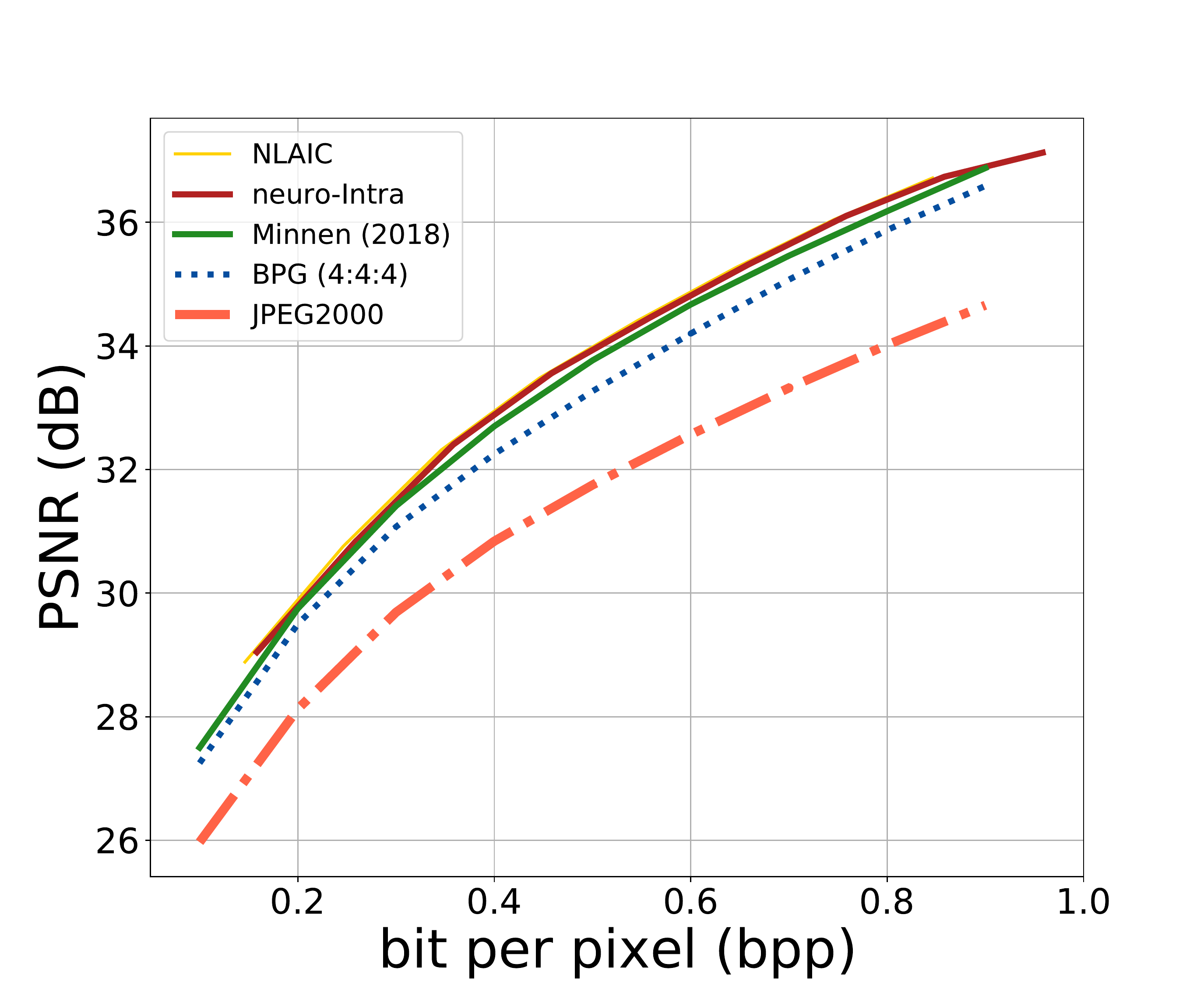}
     \caption{{\bf Efficiency of neuro-Intra.}  PSNR vs. rate performance of neuro-Intra in comparison to  NLAIC~\cite{chen2019neural}, Minnen (2018)~\cite{minnen2018joint}, BPG (4:4:4) and JPEG2000. Note that the curves for neuro-Intra and NLAIC overlap. }
     \label{fig:comp_of_universe_quantization}
\end{figure}

To overcome the non-differentiability of the quantization operation, quantization is simulated by adding uniform noise in~\cite{balle2018variational}. However, such noise augmentation is not exactly consistent with the rounding in inference, yielding the performance loss as reported by~\cite{choi2019variable}. Thus, we apply the universal quantization (UQ)~\cite{choi2019variable} in neuro-Intra, i.e.,: 
\begin{equation}
  {\hat x} = \mathbb{R}\left(x+u\right)-u,
  \label{eq:universal_quantization}
\end{equation}
where $\hat x$ is a quantized symbol and $u$ represents the random uniform variable ranging from $-\frac{1}{2}$ to $\frac{1}{2}$. Statistically, we define the gradient as 1 for back propagation since UQ can be approximated as a linear function. Such UQ  \eqref{eq:universal_quantization} is used for neuro-Motion and neuro-Res as well. When applying to common Kodak dataset, neuro-Intra achieved similar performance as NLAIC~\cite{chen2019neural}, outperforming Minnen (2018)~\cite{minnen2018joint}, BPG (4:4:4) and JPEG2000, as shown in Fig.~\ref{fig:comp_of_universe_quantization}.


\begin{figure*}[t]
\centering
\subfloat[]{\includegraphics[scale=0.19]{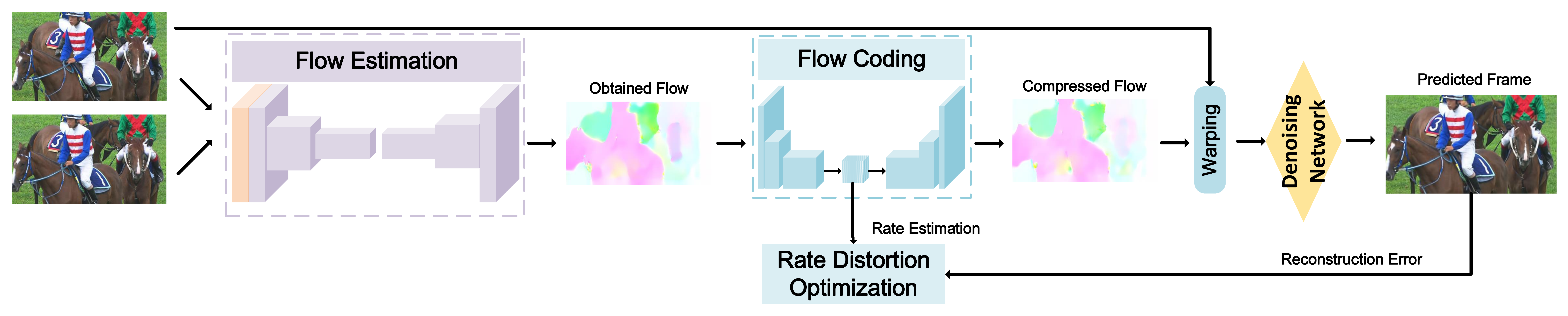} \label{sfig:two_stage}}\\
\subfloat[]{\includegraphics[scale=0.19]{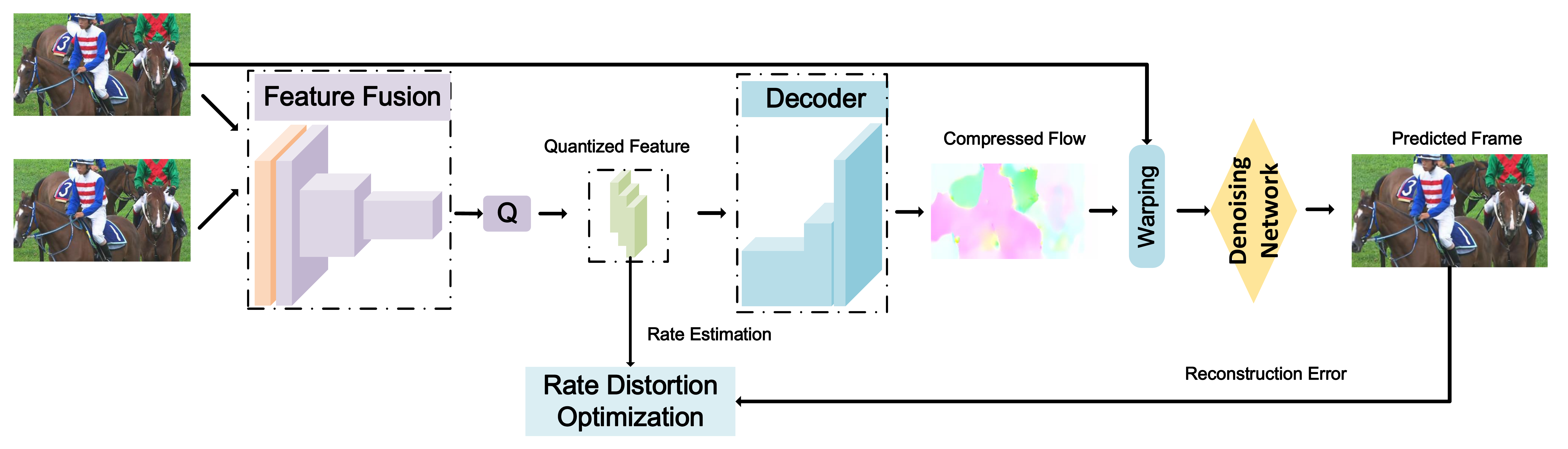} \label{sfig:one_stage}}\\
\subfloat[]{\includegraphics[scale=0.19]{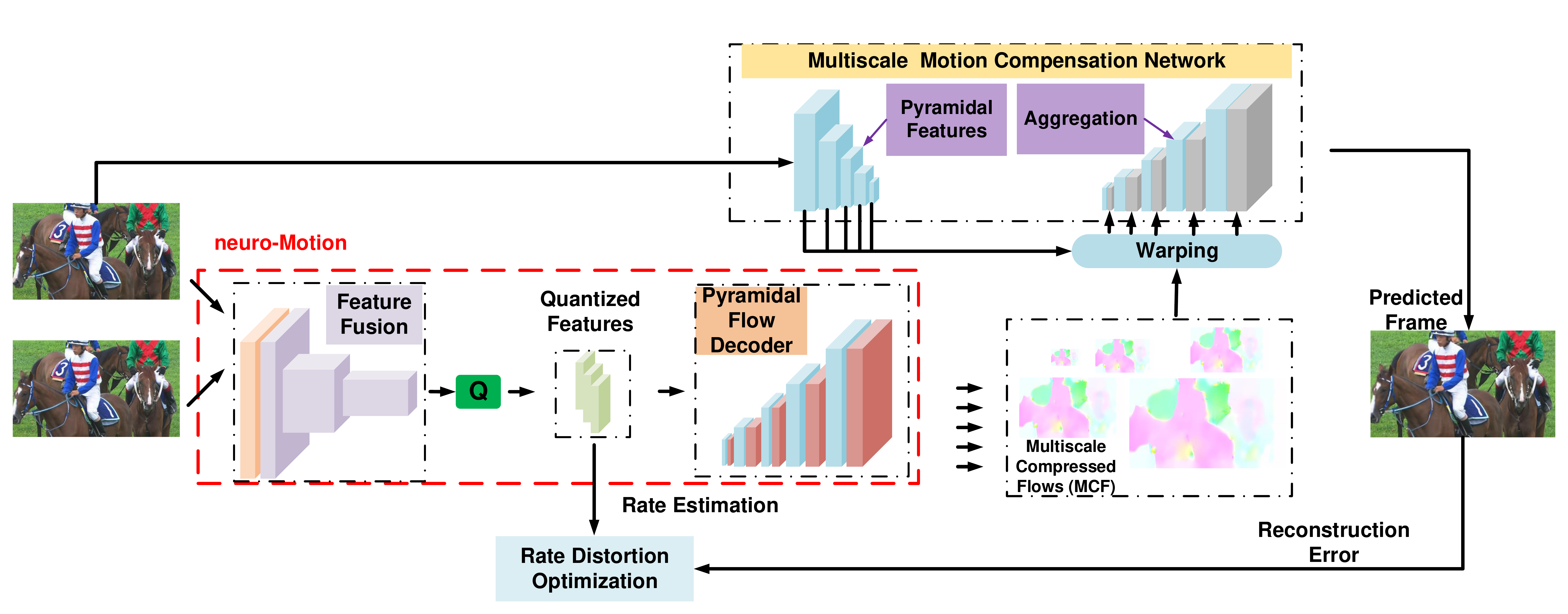} \label{sfig:multi-scale}}
\caption{{\bf Motion Estimation and Compensation. } (a) Two-stage single-scale motion coding and compensation approach using a pre-trained flow network (with explicit raw flow) and  a cascaded flow compression network (e.g.,~\cite{Lu_2019_CVPR}); (b)  One-stage unsupervised motion generation and compensation approach with implicit flow represented by quantized features that will be decoded into a single scale motion field for motion compensation (warping) (e.g., \cite{liu2019learned}); (c) One-stage neuro-Motion with MS-MCN uses a pyramidal flow decoder to synthesize the multiscale compressed flows (MCFs) that are used in a multiscale motion compensation network for generating predicted frames.}
\label{fig:flow_coding}
\end{figure*}

 \subsection{Neural Motion Coding and Compensation} \label{sec:inter_coding}
Inter-frame coding plays a vital role in video coding. The key is how to efficiently represent motion in a compact format for effective compensation. In comparison to the pixel-domain block-based motion estimation and compensation in conventional video coding, we rely on the optical flow to accurately capture the temporal information for {\it motion compensation}.


 \subsubsection{Single-scale Motion Generation and Compensation } \label{ssec:ss_mrc}

Deep learning has spawned several optical flow estimation methods in either supervised or unsupervised way, such as FlowNet2~\cite{ilg2017flownet}, PWC-net~\cite{sun2018pwc}, etc. They mostly focus on acquiring high-precision optical flow between two consecutive (uncompressed) frames  without any compression rate constraint. In video coder, it, however, is more challenging to derive robust flow where bitrate is often limited, and the reference frame is usually lossy encoded with inherent compression noises.

One approach applies a two-stage method shown in Fig.~\ref{sfig:two_stage}, which is utilized by DVC~\cite{Lu_2019_CVPR}. Such two-stage scheme first uses a pre-trained lossless flow generator (e.g., FlowNet2) for explicit flow derivation, and then cascades an auto-encoder to encode the flow. RDO is examined when compressing the optical flow to balance the bits allocation and reconstruction distortion.  Either separable or joint optimization of flow derivation and compression can be applied for such two-stage approach.



We propose an alternative one-stage framework, shown in Fig.~\ref{sfig:one_stage}, which is first presented in~\cite{liu2019learned}. It directly transforms concatenated two frames (e.g., one is the reference from past, and one is current frame) into quantized temporal features that represent the inter-frame motion. These quantized features are decoded into compressed optical flow in an unsupervised way for frame compensation via warping. Such one-stage  scheme  does not require any pre-trained flow network such as FlowNet2 or PWC-net to generate the optical flow explicitly. It allows us to quantize the motion features rather than the optical flows and train the motion feature encoder and decoder together with explicit consideration of quantization and rate constraint. Note that the motion features are generated from the main motion encoder, and hyper encoder is used to generate hyper features for motion features in the VAE model in Fig.~\ref{fig:bio_video_coder}.


{{Loop filters, or quality enhancement networks, can be used in both two-stage and one-stage methods to enhance the quality of flow-warped frame, for improved quality of predicted frame, as shown in Fig.~\ref{sfig:two_stage} and~\ref{sfig:one_stage}. This is mainly because linear interpolation-based backward warping would inevitably introduce artifacts in reconstruction, especially for  cases having flow estimation errors~\cite{jaderberg2015spatial}. A cascaded quality enhancement  model can be devised to alleviate such artifacts and improve the reconstruction quality. End-to-end learning can be further applied to the entire framework.}} 



 \subsubsection{Multiscale Motion Generation and Compensation} \label{ssec:ms_mrc}
 We further extend our earlier work~\cite{liu2019learned} to multiscale motion generation and compensation, for better inter-frame prediction. 
 The neuro-Motion is modified for multiscale motion generation, where the main encoder is used for feature fusion, but we replace the main decoder by  a {\it pyramidal flow decoder}, which generates the {Multiscale Compressed optical Flows} (MCFs). MCFs  will be processed together with the reference frame, using a {\it multiscale  motion compensation} network (MS-MCN) to obtain the predicted frame efficiently, as shown in Fig.~\ref{sfig:multi-scale}. 
 {The overall pipeline has four key steps:}

\begin{figure}[t]
     \centering
         \includegraphics[scale=0.27]{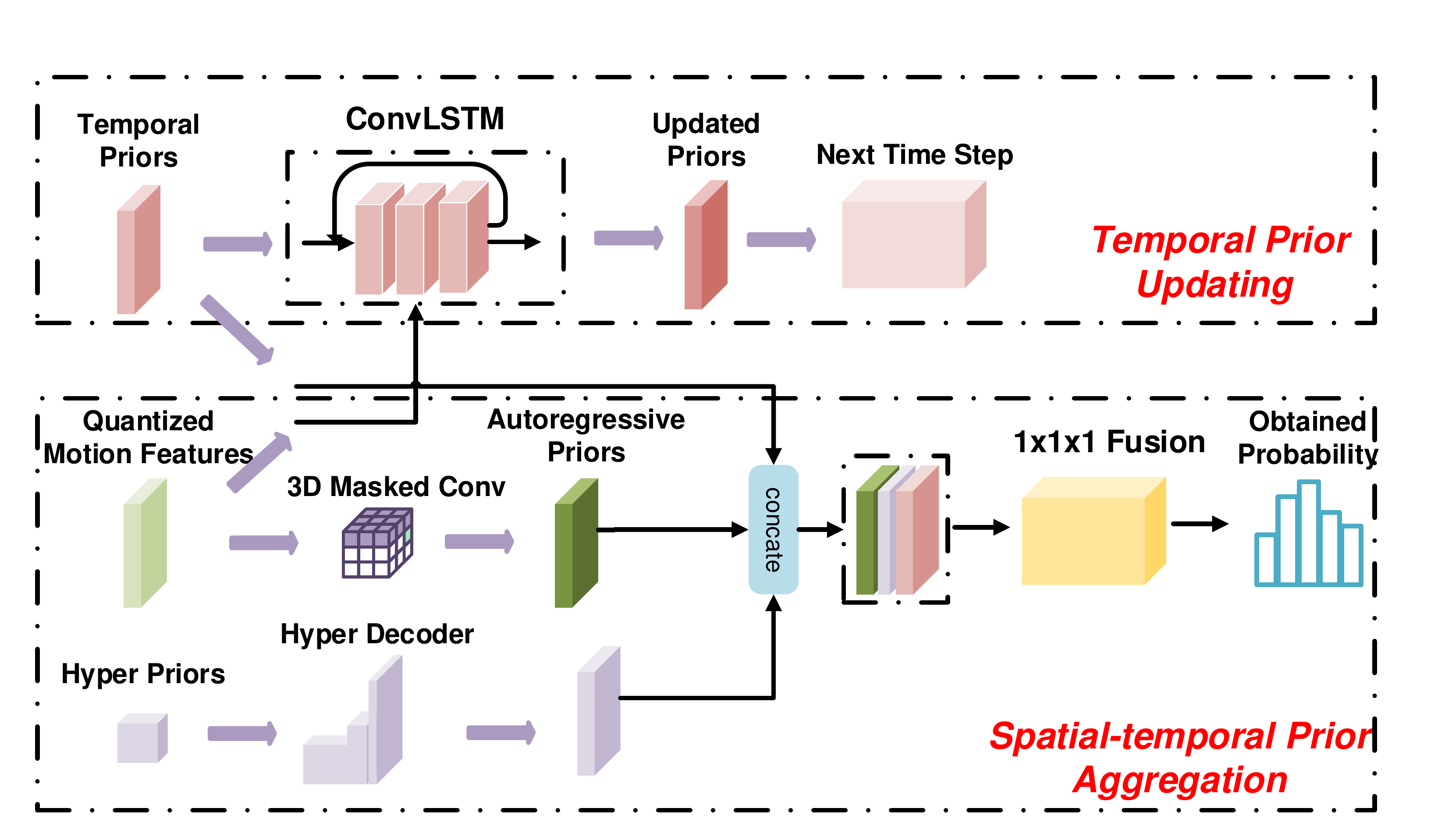}
     \caption{{\bf Context-Adaptive Modeling Using Joint Spatio-temporal and Hyper Priors.}  All priors are fused in PA to provide estimates of the probability distribution parameters. }
     \label{fig:entropy_model}
\end{figure}

\begin{itemize}
    \item  {{\bf Step 0:  Pretrain  the neuro-Motion using $\ell_1$ loss for MCFs generation.} This part involves the implicit motion feature derivation in encoder, and multiscale (compressed) flow decoding in decoder.  
    
    For the former part, we use the main encoder of the VAE (see Fig.~\ref{fig:bio_video_coder}) to extract temporal features for implicit motion representation. It inputs the concatenation of reference frame $\hat{\bf X}_{t-1}$ and current frame ${\bf X}_{t}$ to derive latent motion features. To ensure the accurate MCFs generation, we replace the residual blocks in both VAE encoder and decoder with ResNet-based local attention modules (LAMs)\footnote{Local attention module  only utilizes local features to generate the attention maps compared with NLAM.}. 
    Here, quantized features have a size of $(H/16)\times(W/16)\times 192$ with $H, W$ denoting respective height and width of the original frame.

    For the pyramidal flow decoder, we plug additional convolutional layers at each scale after the LAM to produce resolution-dependent flows $\vec{f}^s_d, s = 0, 1, \ldots, 4$ to capture the multiscale motion fields. The  pyramid flow decoder is first trained in an unsupervised way using the loss:
\begin{align}
  L = {\sum\nolimits^{n=4}_{s=0} }{\alpha}_s\cdot{\mathbb{D}_1(\hat{\bf X}^s_t,{{\bf X}^s_t})} ,
   \label{multi_scale_oss} 
\end{align}
with $\mathbb{D}_1$ as $\ell_1$ loss. ${\alpha}_s$s are scale-dependent weighting coefficients which is set as $\alpha_s = 4^s$ empirically. Here, {$s = 0$} refers to the finest resoltion. $\hat{\bf X}^s_t$ is obtained by backward warping in each scale:
\begin{align}
  \hat{\bf X}^s_t & = {\sf warping}(\hat{\bf X}^s_{t-1},\vec{f}^s_d ). \label{eq:warping} 
\end{align}

Multiscale labels ${\bf X}^s_t$, and multiscale references ${\hat{\bf X}^s_{t-1}}$ are generated from the current and reference frames respectively using average pooling with stride 2 for both vertical and horizontal dimensions.}

\begin{figure}[t]
     \centering
     \subfloat[]{\includegraphics[scale=0.28]{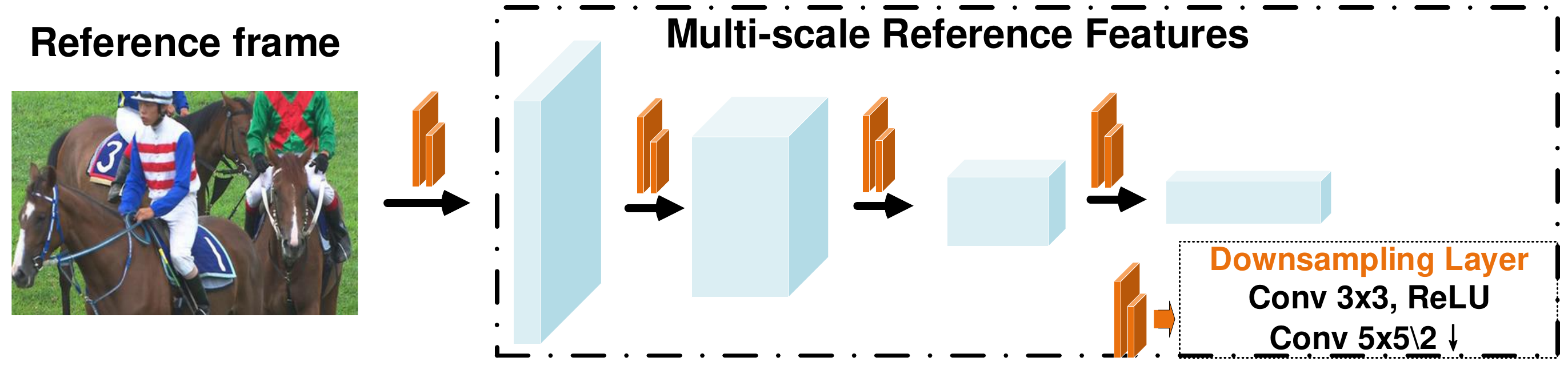}\label{sfig:pyramid}}\\
     \subfloat[]{\includegraphics[scale=0.25]{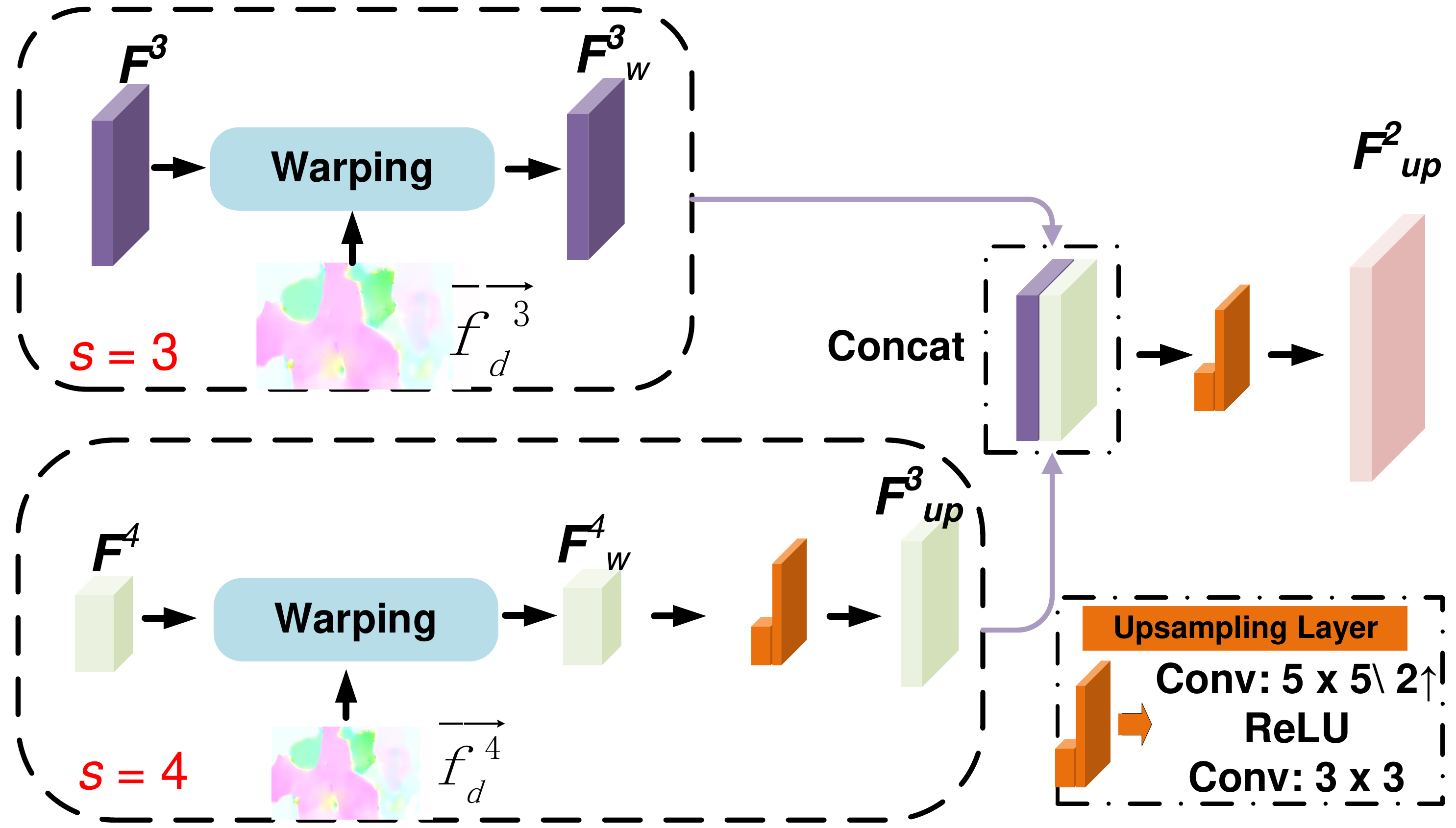} \label{sfig:refinement}}
     \caption{{\bf Multiscale Motion Compensation.} (a) A pyramidal decomposition is applied on reference frame to generate multi-scale features. It consecutively uses a 3$\times$3 convolution, ReLU activation and a 5$\times$5 convolution with stride 2. (b) Feature Aggregation between consecutive scales. Each upsampling layer consists of a 5$\times$5 convolution with stride 2, ReLU activation and a 3$\times$3 convolution.}
\end{figure}

\item {\bf Step 1: Fine-tune neuro-Motion via rate-constrained $\ell_1$ loss.} 
We further fine-tune the neuro-Motion with rate constraints, where the bit rate of quantized feature representations will be estimated based on adaptive contexts generated from the hyper features and other spatiotemporal context (to be discussed in Fig.~\ref{fig:entropy_model}).We use a rate-distortion loss as in~\eqref{eq:rdo} where the distortion is the multiscale prediction loss as in~\eqref{multi_scale_oss} and the rate is the estimated entropy of the motion features.

\item  {{\bf Step 2: Pretrain MS-MCN via MCFs.} Using the MCFs (e.g., $\vec{f}^s_d$s) generated by the neuro-Motion from {\bf Step 1}, we will pretrain the {\it multiscale motion compensation network} for motion compensated prediction in the feature domain. It first uses a pyramidal decomposition of the reference frame, shown in Fig.~\ref{sfig:pyramid}, to generate multiscale features $\{{\bf F}^s\}, s =0,1\ldots4$ of the reference frames $\hat{\bf X}_{t-1}$ which are respectively warped with the MCFs at corresponding scale for progressive aggregation. 


Fig.~\ref{sfig:refinement} exemplifies the aggregation at the smallest with $s=4$ and the second smallest scale with $s=3$: we use  $f^4_d$ to warp corresponding ${\bf F}^4$ using \eqref{eq:warping}, leading to the warped representation ${\bf F}^4_{w}$. This ${\bf F}^4_{w}$ is then upsampled to  ${\bf F}^3_{up} = \mathbb{U}({\bf F}^4_{w})$ that has the same resolution as the next scale, and concatenated with ${\bf F}^3_{w}$. These concatenated features are then upsampled again to yield scale 2 features:
\begin{align}
     {\bf F}^2_{up} = \mathbb{U}\left(CAT({\bf F}^3_{w},{\bf F}^3_{up})\right) = \mathbb{U}\left(CAT({\bf F}^3_{w}, \mathbb{U}({\bf F}^4_{w})\right)).
\end{align} Note that ${\bf F}^2_{up}$ will be concatenated with  ${\bf F}^2_{w}$ using similar steps. The upsampling operator $\mathbb{U}()$ is implemented with two convolution layers  as detailed in Fig.~\ref{sfig:refinement}. Eventually, we will apply a fusion layer, consisting of a 1$\times$1 convolution, ReLU and  a 3$\times$3 convolution, on  $CAT(\left({\bf F}^0_{w},{\bf F}^0_{up}\right))$ to obtain the final predicted frame.} Pretraining the MS-MCN is accomplished by minimizing the multiscale motion compensation loss in~\eqref{multi_scale_oss}.


\item  {\bf Step 3: End-to-end Joint Refinement.} In the end, we take pre-trained neuro-Motion and MS-MCN together and perform an end-to-end joint refinement. We use the rate-constrained loss as in {\bf Step 1}, and use the reconstructed intra frame as the reference frame.
We resort to the ``pre-training and joint-refinement'' strategy since applying the joint training directly makes the model extremely unstable and hard to converge as revealed experimentally. 

\end{itemize}


 \begin{table}[b]
  \centering
  \caption{Comparative Studies of MS-MCN and SS-MCN Using the PSNRs of Predicted Frame.}
  
  \begin{tabular}{c|c|c|c|c}
    \hline
    \multirow{2}{*}{\textbf{Sequences}}&\multicolumn{2}{c|}{\textbf{High Bit Rate}}&\multicolumn{2}{c}{\textbf{Low Bit Rate}}\\
     
    \cline{2-5}
      & MS-MCN & SS-MCN & MS-MCN & SS-MCN \\
     \hline
    BasketballPass &\bf 32.94& 29.34 &\bf 30.52  &30.01\\
    \hline
    RaceHorses &\bf 27.48& 25.08 &\bf 26.12 & 24.41\\
    \hline
    PartyScene &\bf 27.01& 26.43 &\bf 25.55 &25.27\\
    \hline
    BQMall & \bf 33.09& 31.35 & \bf 30.76 & 30.33 \\
    
    \hline
    vidyo1 &\bf38.81 & 37.14 &\bf 36.34  & 36.30\\ 
    \hline
    vidyo4 &\bf 38.86& 36.60 &\bf 36.40 & 36.33\\
    \hline
    Average &\bf 33.03& 30.99 &\bf 30.94 & 30.44\\
    \hline
  \end{tabular}
  \label{tab:compare_prediction}
\end{table}

Table~\ref{tab:compare_prediction} reports the consistent PSNR gains offered by the MS-MCN in Sec.~\ref{ssec:ms_mrc}, in comparison to the SS-MCN~\cite{liu2019learned} in Sec.~\ref{ssec:ss_mrc}, revealing the efficiency of MS-MCN for accurately motion compensated prediction. At high bit rate scenario, almost 2 dB gain is observed; MS-MCN keeps superior efficiency even with challenging temporal motions or occlusions in ``BasketballPass'' content. At low bit rates, gains are reduced but still noticeable. It shows that the PNSR gains depend on how much motion information are actually compressed. The more information (e.g., high bit rate) comes with larger PSNR gains, and vice versa.

\subsubsection{Context-Adaptive Flow Coding} \label{ssecion:flow_coding_contexti}

It is well recognized that motion fields present high correlations, both spatially and temporally. To exploit such correlation, Ripple et al.~\cite{rippel2019learned} proposed to code the flow differences in feature domain.
Here, we propose to exploit this correlation in the context modeling for the entropy coding of the motion features. Specifically, we develop  a  joint spatiotemporal and hyper prior-based context-adaptive model shown in Fig.~\ref{fig:entropy_model}. This is implemented in PA engine of Fig.~\ref{fig:bio_video_coder} for neuro-Motion.

The proposed PA engine for neuro-Motion consists of  a {\it spatio-temporal-hyper aggregation module} (STHAM)  and a {\it temporal updating module} (TUM), shown in Fig.~\ref{fig:entropy_model}. At timestamp $t$, STHAM accumulates all the accessible priors and estimate the mean and standard deviation of the assumed Gaussian distribution for each new quantized motion feature $\mathscr{\hat F_i}$ using:
\begin{equation}
  (\mu_{\mathscr{\hat{F}}},\sigma_{\mathscr{\hat{F}}}) = \mathbb G( {\mathscr{\hat{F}}}_1,..., {\mathscr{\hat{F}}}_{i-1}, \hat{\bf z}_t, {\bf h}_{t-1}), \label{eq:flow_probability}
\end{equation}
Here, $\mathscr{\hat{F}}_i, i =  1, 2, ...$ are elements of quantized latent motion features for the current frame, ${\bf h}_{t-1}$ is consists of temporal priors derived from the motion features preceding the current frame. $\hat z_t$ includes hyper priors of the quantized motion features.  These features are concatenated and fused using stacked 1x1x1 convolutions. Note that masked convolution is used for the spatial features $\mathscr{\hat F_i}$.  


To generate the temporal priors, TUM is applied to current quantized features $\mathscr{\bf \hat{F}_t}$ recurrently using a standard ConvLSTM:
\begin{equation}
  ({\bf h}_t, {\bf c}_t) = {\rm ConvLSTM}({\mathscr{\bf \hat{F}_t}, {\bf h}_{t-1}, {\bf c}_{t-1}}),
  \label{eq:trn_priors}
\end{equation}
where ${\bf h}_t$ are updated temporal priors for the next frame, ${\bf c}_t$ is a memory state to control information flow across multiple time instances (e.g., frames). Note that the STHAM and TUM components are trained with other components in neuro-Motion in {\bf Step 1} and {\bf Step 3} described in Sec.~\ref{ssec:ms_mrc}.




It is worth to point out that leveraging temporal correlation for compact motion representation is also widely explored in traditional video coding approaches. For example, motion vector predictions from spatial and temporal co-located neighbors are standardized in HEVC, by which only motion vector differences (after prediction) are encoded. Here, instead of coding the flow feature differences, we use the flow features in the past to help estimating the probability distribution of the flow features in the current frame.

\subsection{Neural Residual Coding} \label{ssec:residual_coding}

Inter-frame residual coding is another significant module contributing to the overall system efficiency. It is used to compress the temporal prediction error pixels.  It affects the efficiency for next frame prediction since errors usually propagate temporally. 

Here we use the VAE architecture in Fig.~\ref{fig:bio_video_coder} for encoding the  residual ${\bf r}_t$. The rate-constrained loss is used:  
\begin{equation}
  L = \lambda\cdot\mathbb{D}_2\left({\bf X}_t,({\bf X}^p_t+{\hat{\bf r}_t})\right) + R,
  \label{eq:residual_generation}
\end{equation} where $\mathbb{D}_2$ is the $\ell_2$ loss between a residual compensated frame ${\bf X}^p_t+{\hat{\bf r}_t}$ and ${\bf X}_t$. neuro-Res will be first pretrained  using the predicted frames by the pretrained neuro-Motion and MS-MCN, and a loss function in~\eqref{eq:residual_generation} where the rate $R$  only consider the bits for residual. Then we refine it jointly with neuro-Motion and MS-MCN, using a loss where $R$ considers the bits for both motion and residual with two frames. 

\subsection{Training Strategy}
{\bf Progressive Training.} 
Training all NVC components directly is difficult and unreliable, because these modules are interdependent with each other. For example, inter prediction depends on the former reconstructed frame and meanwhile a better predicted frame will reduce the residual energy. In our model development, we first train the neuro-Intra for multiple bit rates, using multiple $\lambda$ values; Then we pretrain and fine-tune the neuro-Motion and MS-MCN through {\bf Step 1-3} described in Sec.~\ref{ssec:ms_mrc}, to minimize a training loss that consider both the prediction error and the rate for motion features. We also pretrain neuro-Res using the predicted frames generated by the trained neuro-Motion and MS-MCN so far, with  a loss that considers both bit rate of residual and the final reconstruction of the current frame. Finally, we further refine  all the modules for inter-coding, including neuro-Motion, MS-MCN and neoro-Res, using a training loss including the reconstruction error and the total rate for the motion features and residual features.   For each target bit rate, we set the  $\lambda$ for training the inter-coding model to be proportional to the $\lambda$ used for training the neuro-Intra module.


 {\bf Training with Multi-frame Loss.} One way to train the inter-coding model (including neuro-Motion, MS-MCN and neuro-Res modules)  is to use only two frames as a training sample and use the neuro-Intra module to code the first frame, and use the inter-coding model to code the second frame. This training approach  can only learn  the intra-to-inter variations, yielding instability and poor reconstruction for the succeeding inter frames distant from the first intra frame. To overcome the quality degradation in future inter-frames, we adopt a multi-frame training strategy.
 
 We first pretrain the inter-coding model using pairs of two successive frames as training samples, where the first frame is encoded and decoded using the neuro-Intra. This step follows the progressive training procedure. We then refine the inter-coding model by using groups of four successive frames as training samples, where the first frame is encoded and decoded by a pretrained neuro-Intra, and each following frame is coded using the inter-coding model and the decoded frame is then used recursively as the reference frame for coding the next frame. We use a loss function that considers the sum of the distortions (in MSE or negative MS-SSIM)  in all three P-frames, and the sum of the rate for these frames, so that the model update considers the impact of the propagation of P-frame reconstruction errors. Note that once the inter-coding model is trained, it is applied to all inter frames in a GOP during testing. Although it is possible to improve the performance by training a different inter-coding model for each possible distance between a P-frame and the intra-frame, we choose to train a single model that is used repeatedly for all P-frames, to reduce the implementation complexity.

%% file: exp.tex
\section{Experimental Studies} \label{sec:exp}
\begin{figure*}[h]
\centering
\subfloat[]{\includegraphics[scale=0.26]{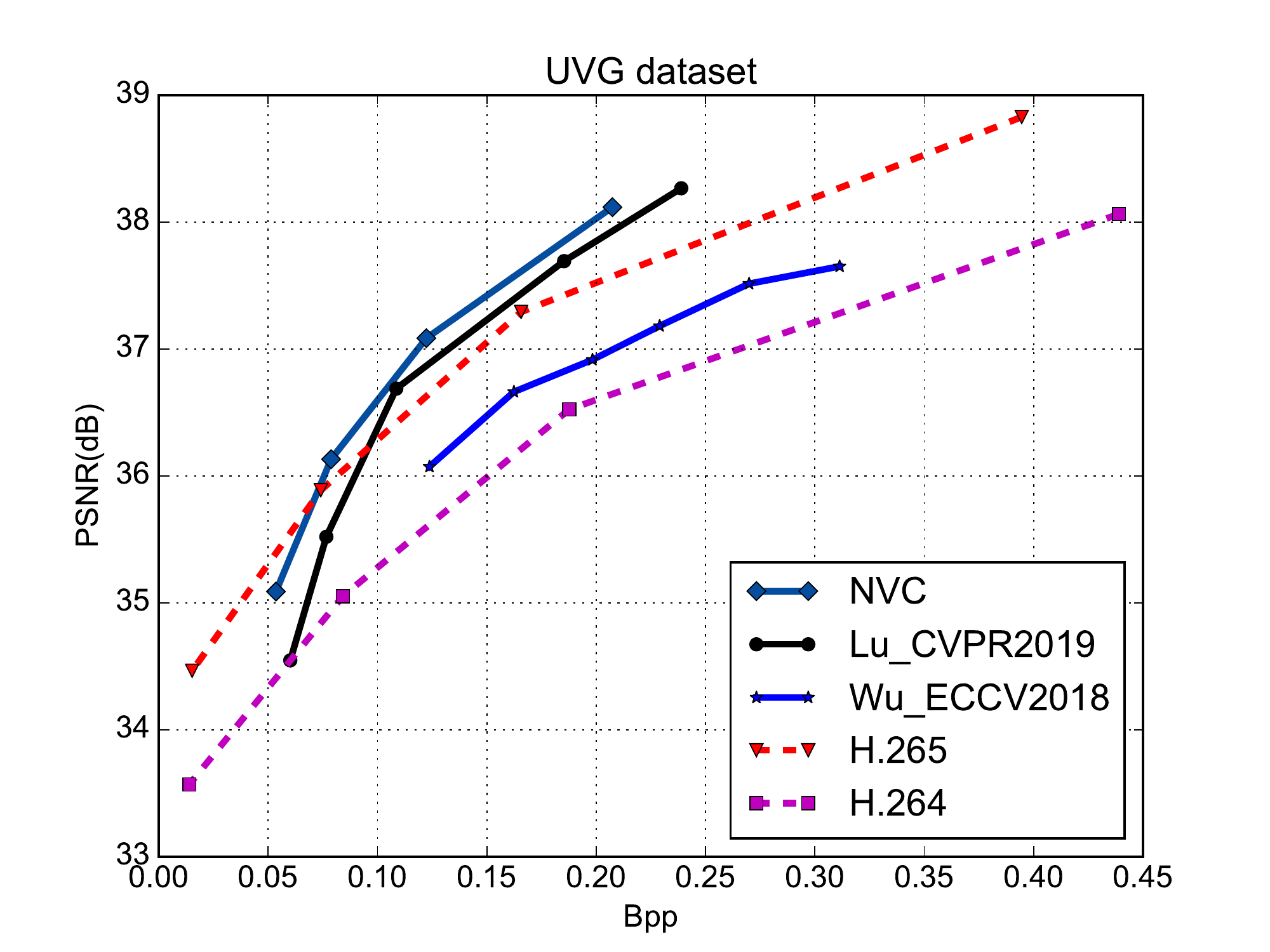}}
\subfloat[]{\includegraphics[scale=0.26]{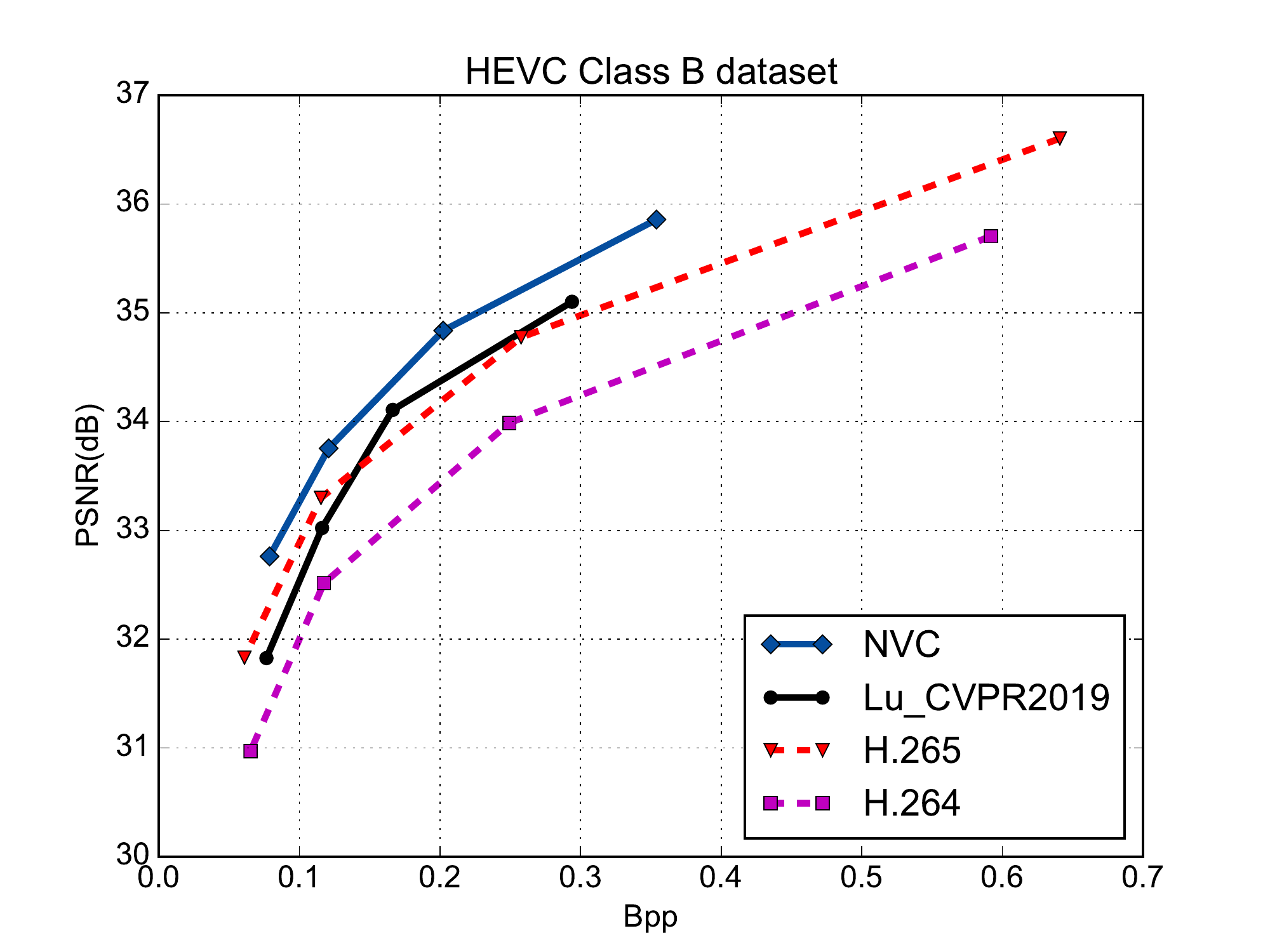}}
\subfloat[]{\includegraphics[scale=0.26]{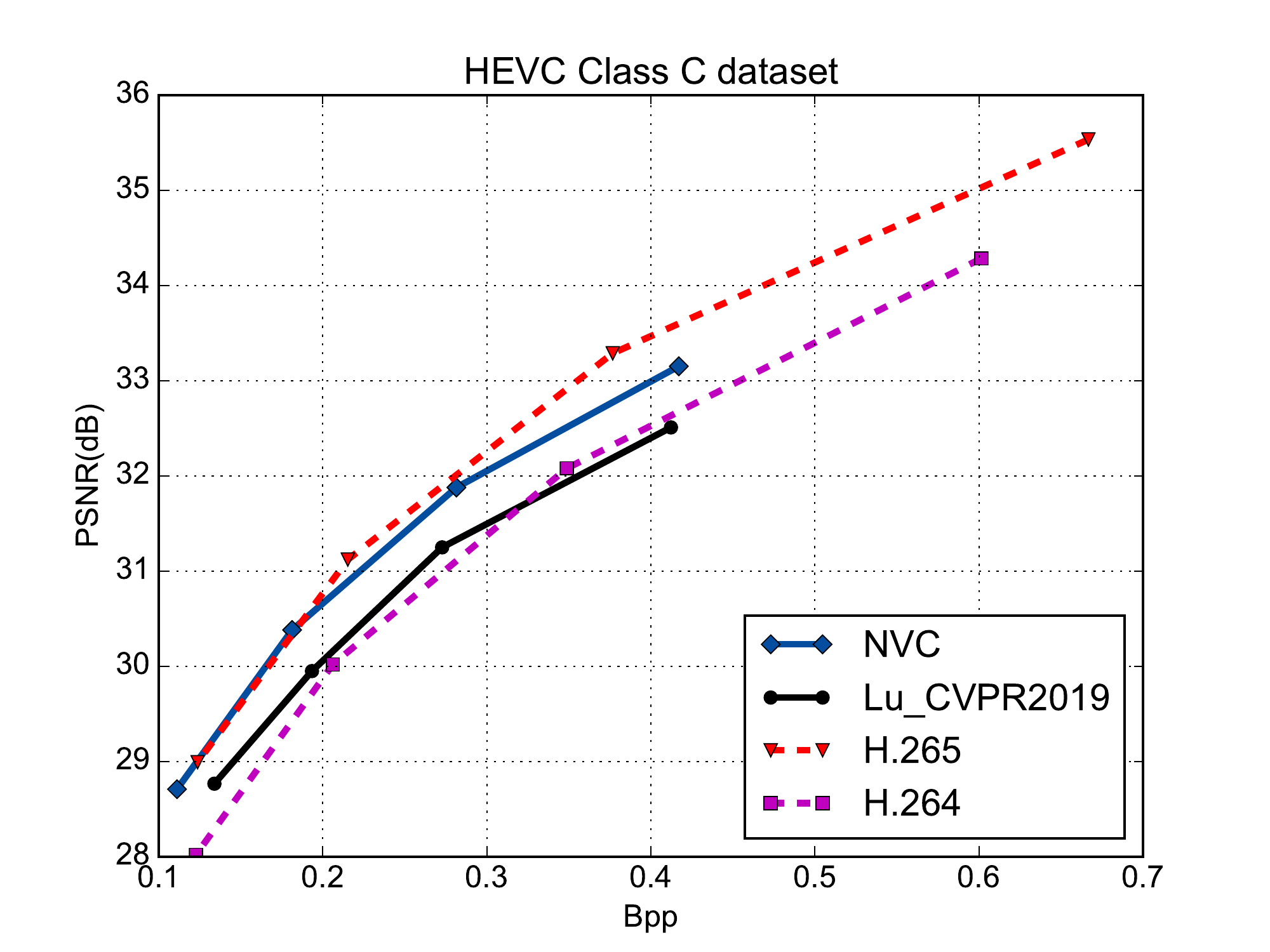}}\\
\subfloat[]{\includegraphics[scale=0.26]{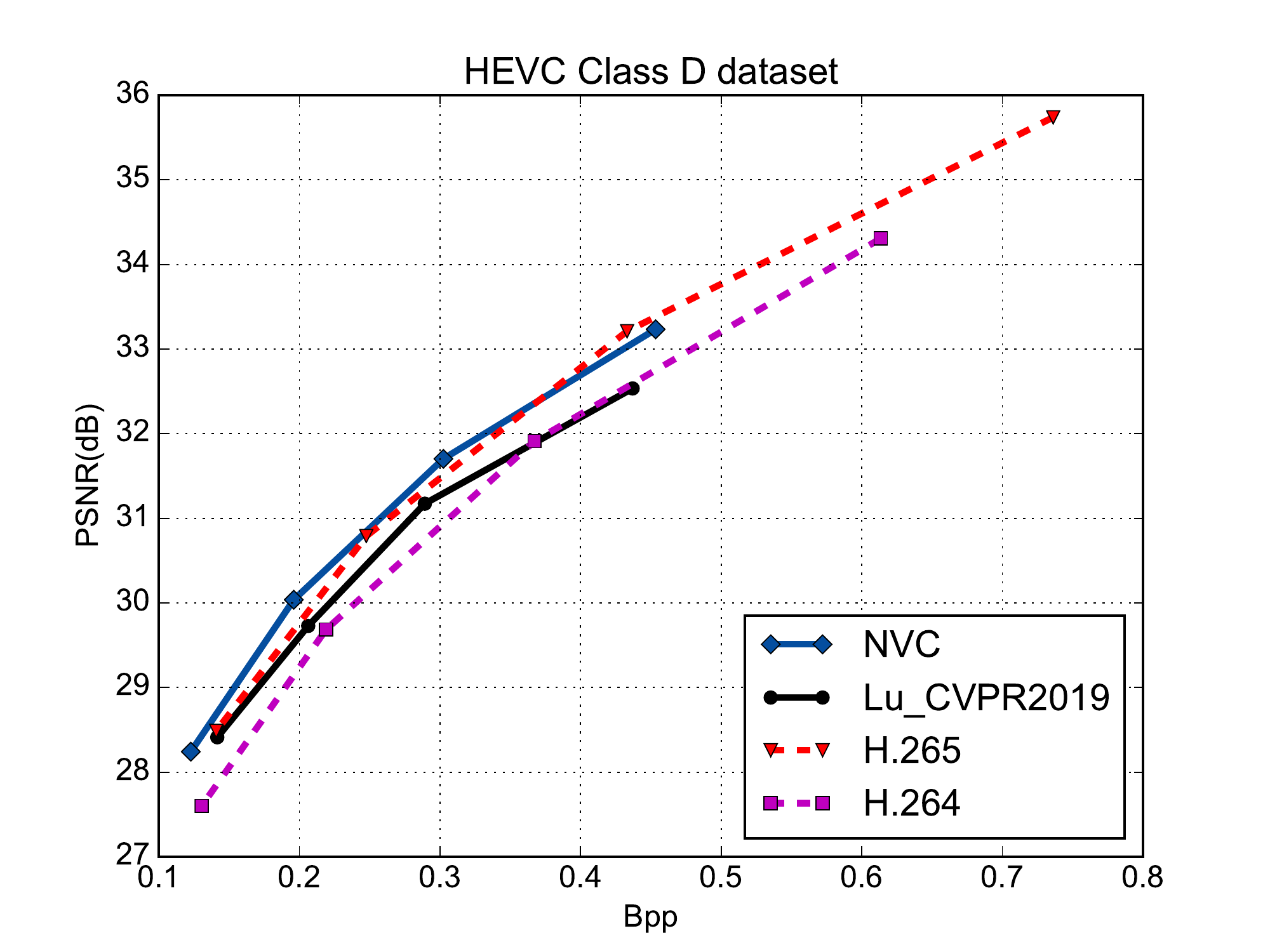}}
\subfloat[]{\includegraphics[scale=0.26]{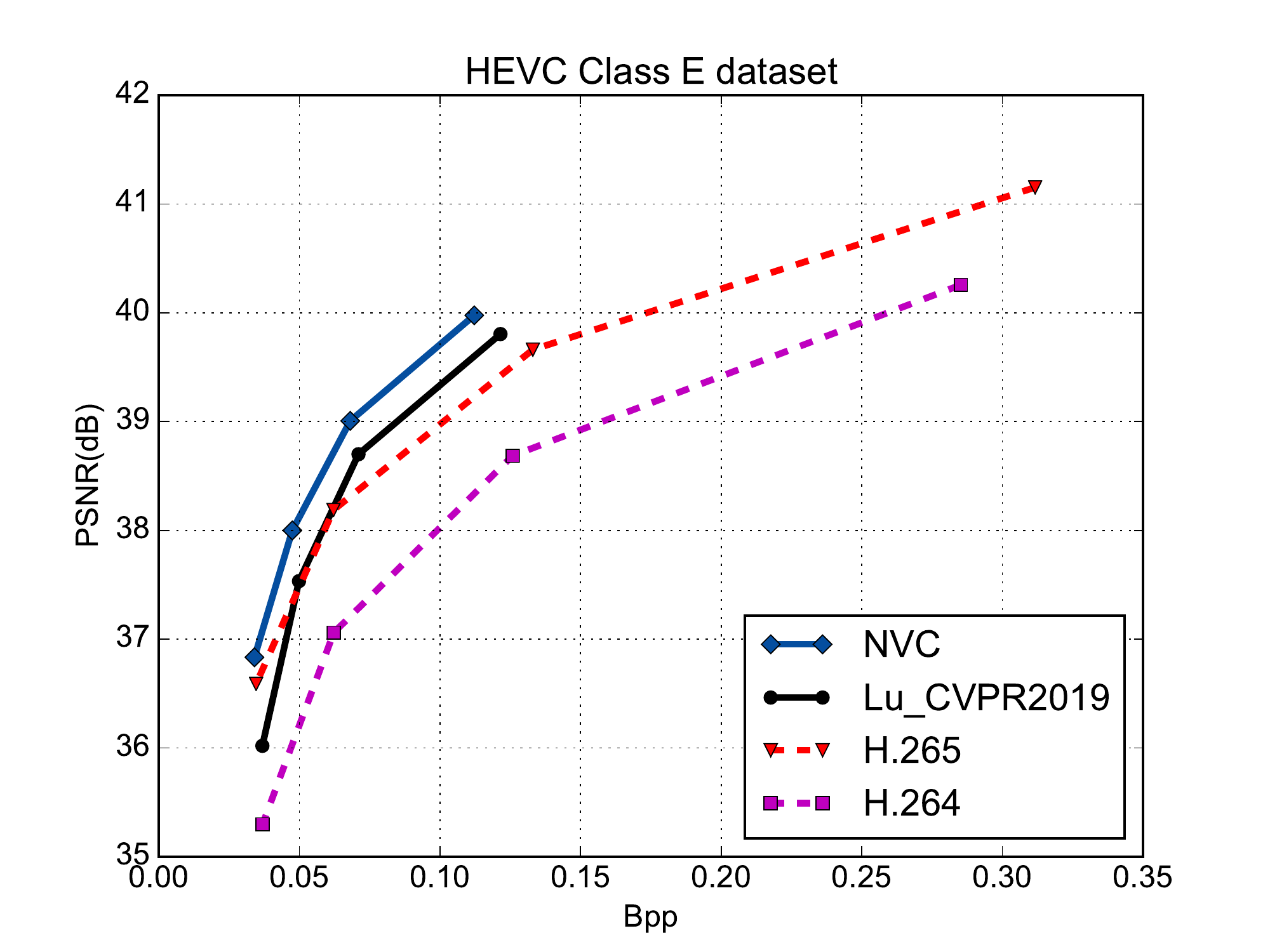}}
\subfloat[]{\includegraphics[scale=0.26]{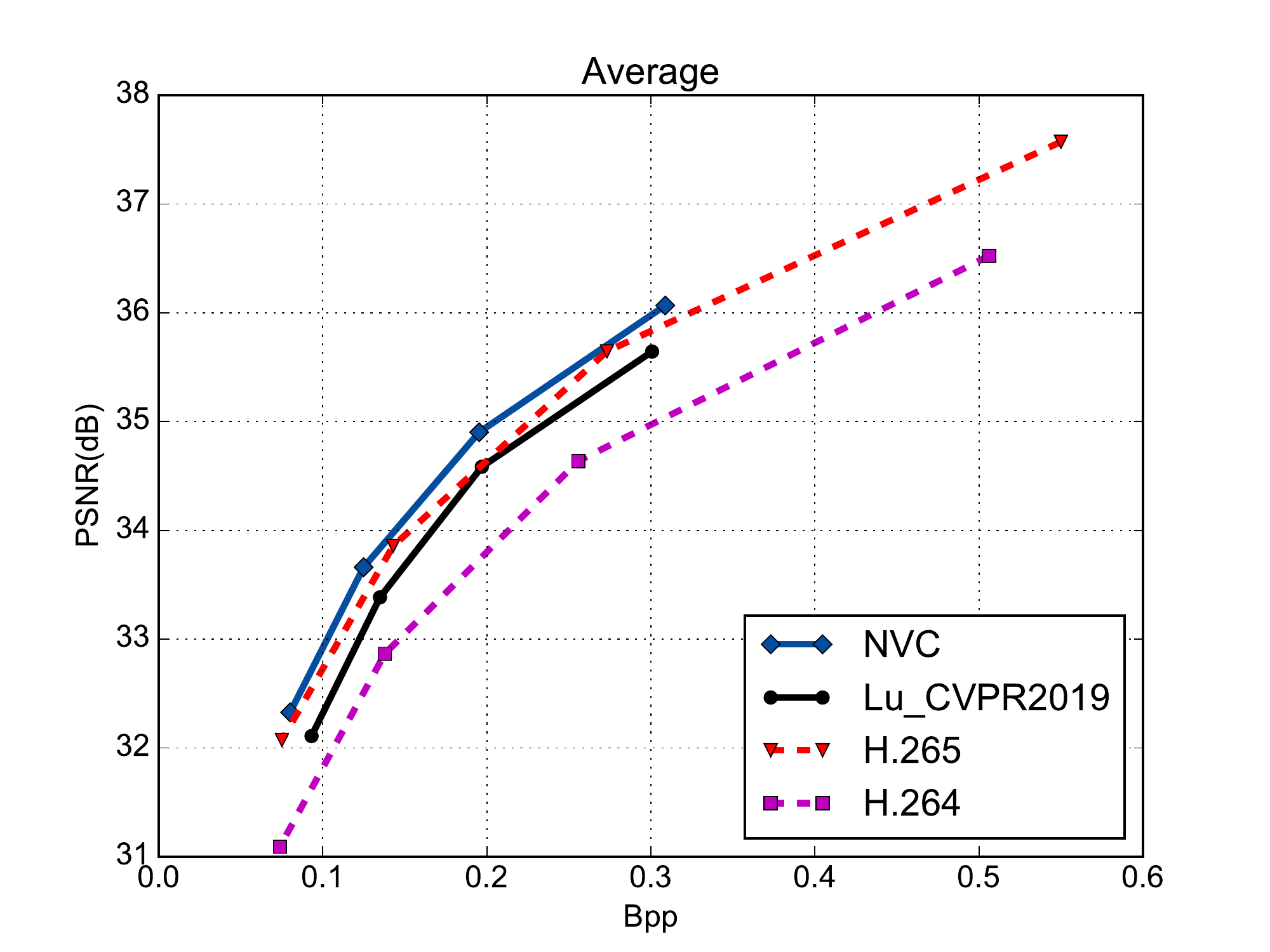}}
\caption{{\bf PSNR vs. Rate Performance.}
}
\label{rd_curve_psnr}
\end{figure*}
\subsection{Datasets and Hyperparameters for Model Training }

{\bf Datasets.} The neuro-Intra is trained on COCO~\cite{lin2014microsoft} and CLIC~\cite{clic} with training samples randomly cropped into 256$\times$256$\times$3. The neuro-Motion, MS-MCN and neuro-Res are joint trained using Vimeo 90k~\cite{xue2019video} with sample size of 192$\times$192$\times$3.

Our NVC is evaluated using the standard HEVC test sequences and ultra video group (UVG) dataset. HEVC test sequences include different classes, covering the contents with a variety of motion, frame rate, resolution and texture; UVG dataset has seven 1080p videos which are often selected for testing video applications.

{\bf Loss Function.} 
We have used either MSE or negative MS-SSIM loss for training the intra-coding and inter-coding modules. For pretraining the neuro-Motion and MS-MCN modules, we use the $\ell_1$ loss on the prediction error instead which is similar to the mean absolute error (MAE) used in traditional motion estimation. We do not use MS-SSIM loss on the predicted frame because MS-SSIM mainly cares about the structure similarity and may ignore the background noise to some extent. Through experiments, we have found that using MS-SSIM loss for neuro-Motion can lead to temporal error accumulation, which not only increases the bits consumption but also potentially leads to unreliable training.

{\bf Hyperparameters and Platform.} The initial learning rate (LR) is set to 10e-4 and is halved for every 10 epochs, and final models are obtained using a LR of  10e-5  for  both  pretraining  and  overall  training.  We  apply the distributed training on 4 GPUs (Titan Xp) for 3 days for each bit rate model.

{\bf Evaluation Criteria.}  We apply the same low-delay coding setting as DVC in~\cite{Lu_2019_CVPR} for our method and traditional H.264/AVC, and HEVC for fair comparison. We encode 100 frames and use GOP of 10 on HEVC test sequences, and 600 frames with GOP of 12 on UVG dataset.  Both PSNR and MS-SSIM results are offered to understand the efficiency of NVC.

{\bf  Generate Models for Different Target Rates.} 
To generate a model for a particular bit rate, we first train the neuro-Intra using a certain  $\lambda$ value, denoted by $\lambda_{\rm intra}$. Then  we train the   inter-frame models (neuro-motion, MS-MCN and neuro-Res) using a proportionally reduced $\lambda$ value, $\lambda_{\rm inter}= \lambda_{intra}/4.$  This simple way of setting the $\lambda_{\rm intra}$ and $\lambda_{\rm inter}$ values yielded good results. We chose not to further optimize $\lambda_{\rm inter}$ for given $\lambda_{\rm intra}$.

\subsection{Performance Comparison}

\begin{figure*}[h]
\centering
\subfloat[]{\includegraphics[scale=0.26]{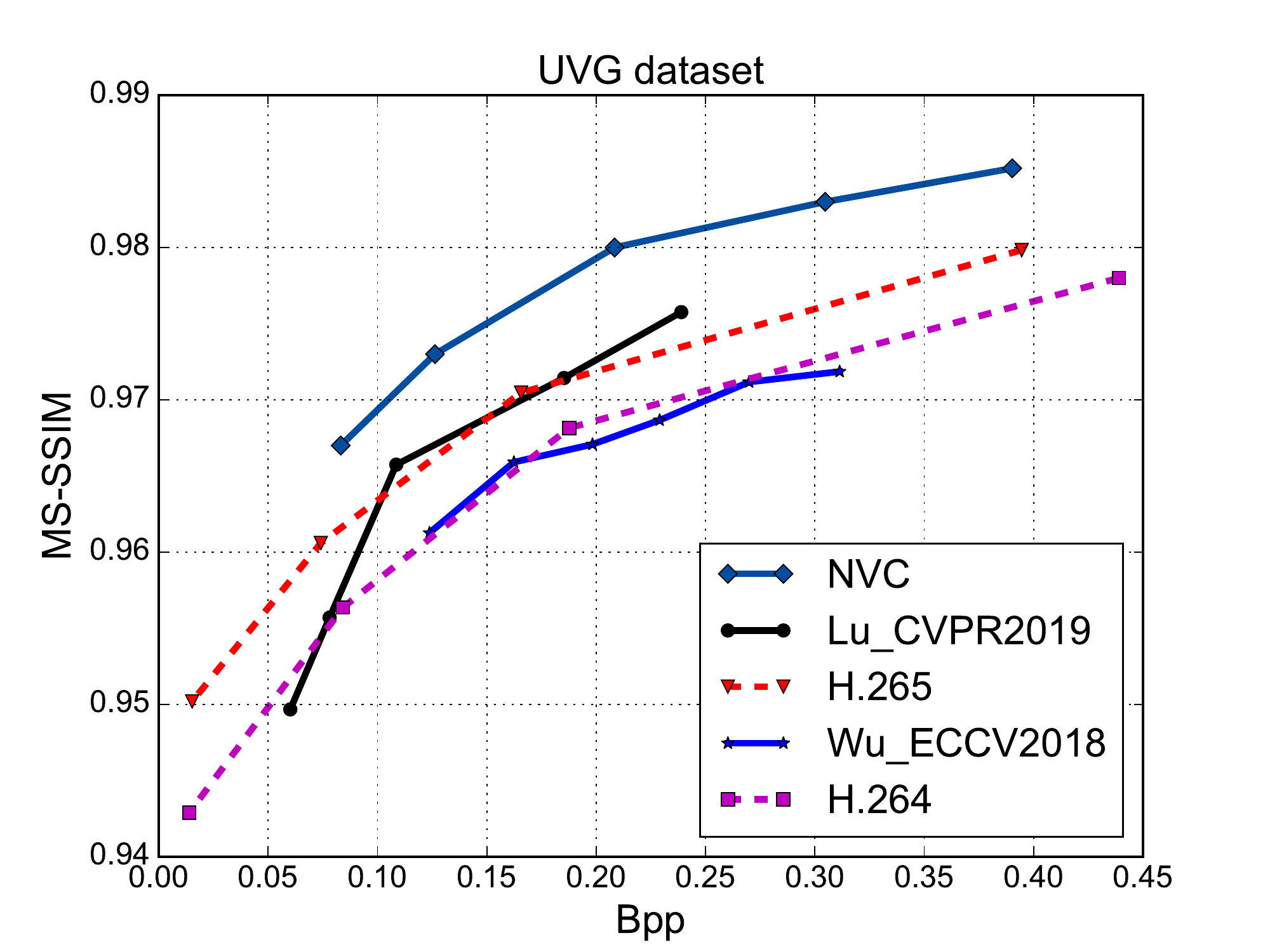}}
\subfloat[]{\includegraphics[scale=0.26]{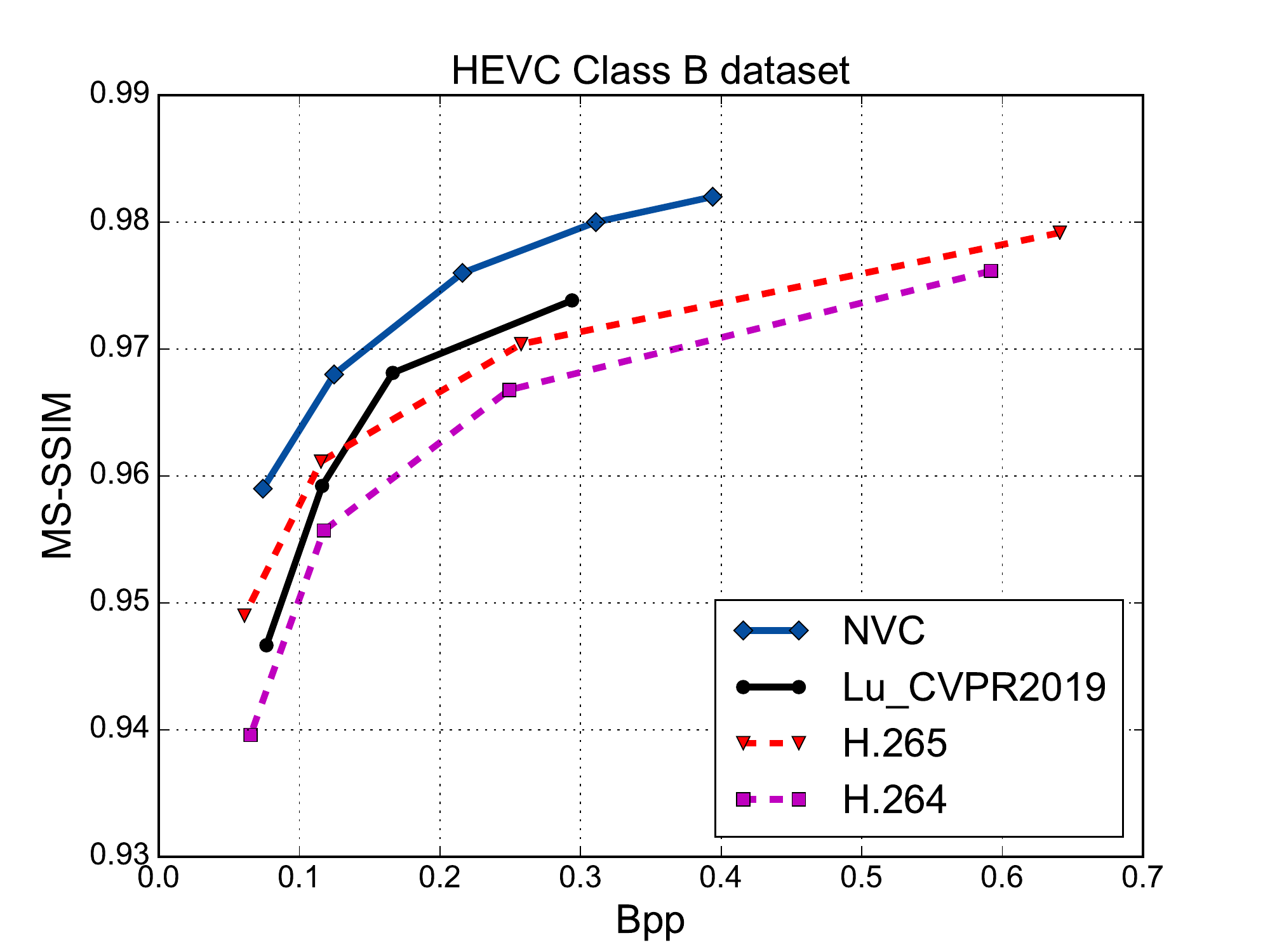}}
\subfloat[]{\includegraphics[scale=0.26]{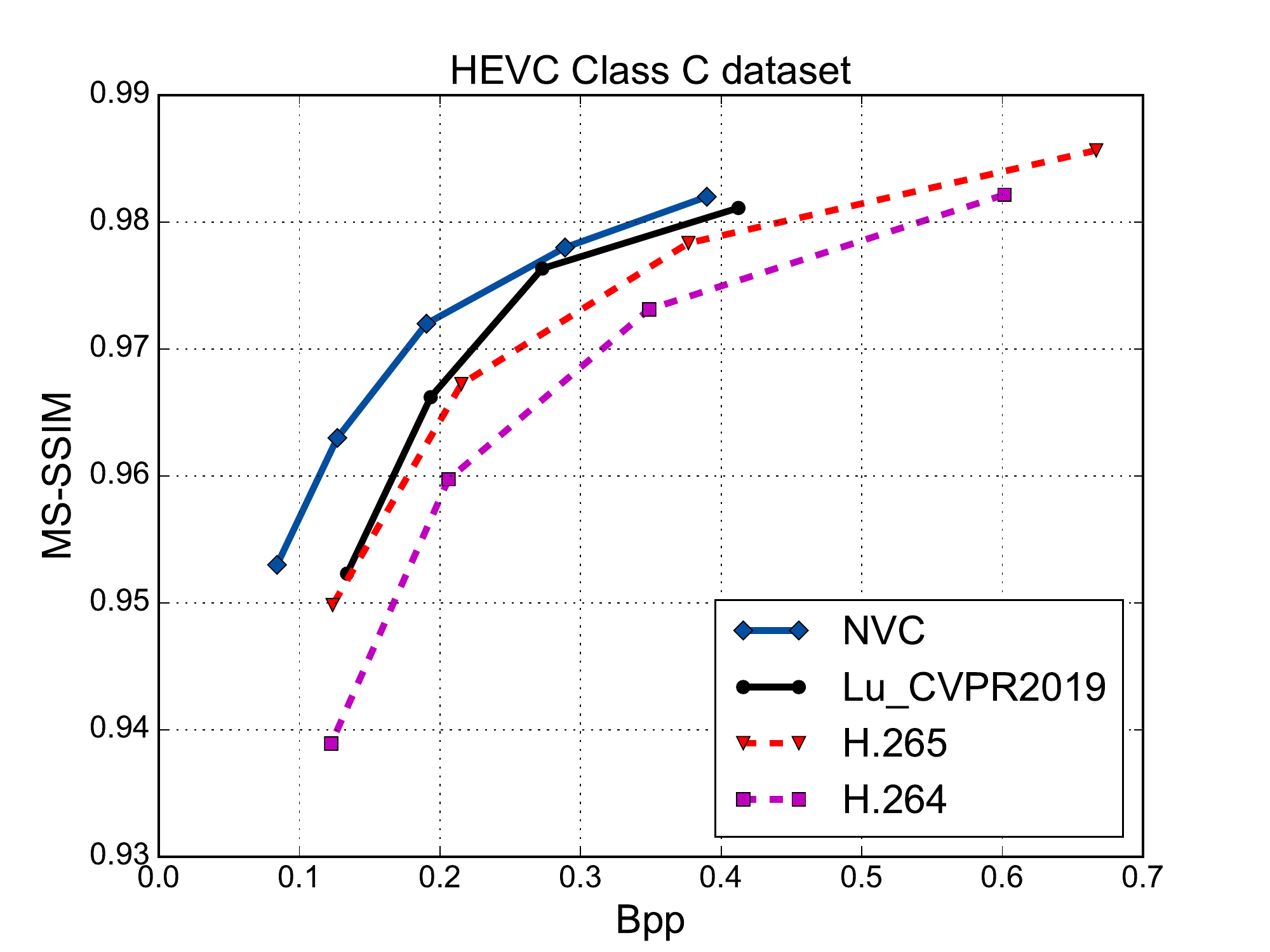}}\\
\subfloat[]{\includegraphics[scale=0.26]{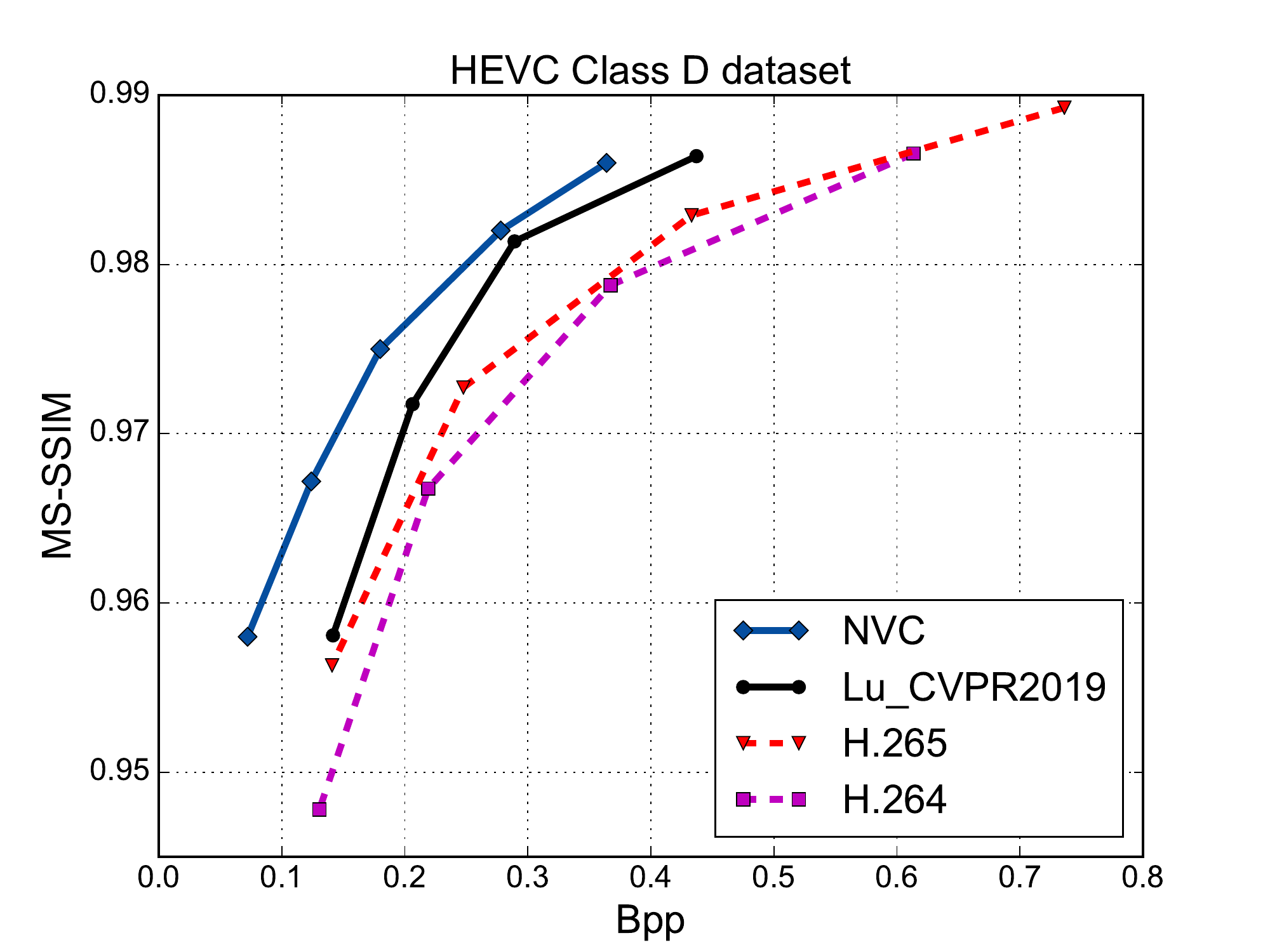}}
\subfloat[]{\includegraphics[scale=0.26]{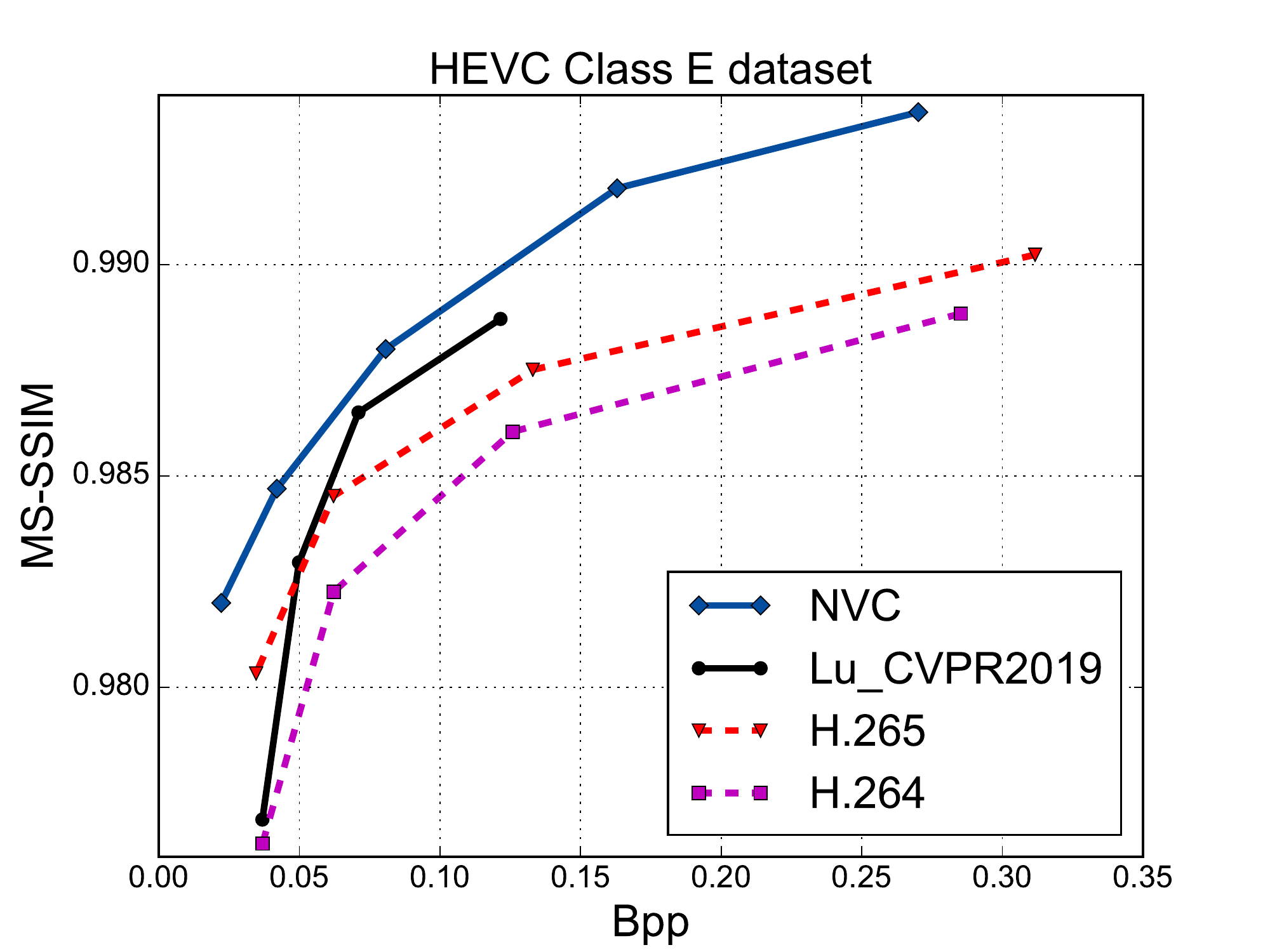}}
\subfloat[]{\includegraphics[scale=0.26]{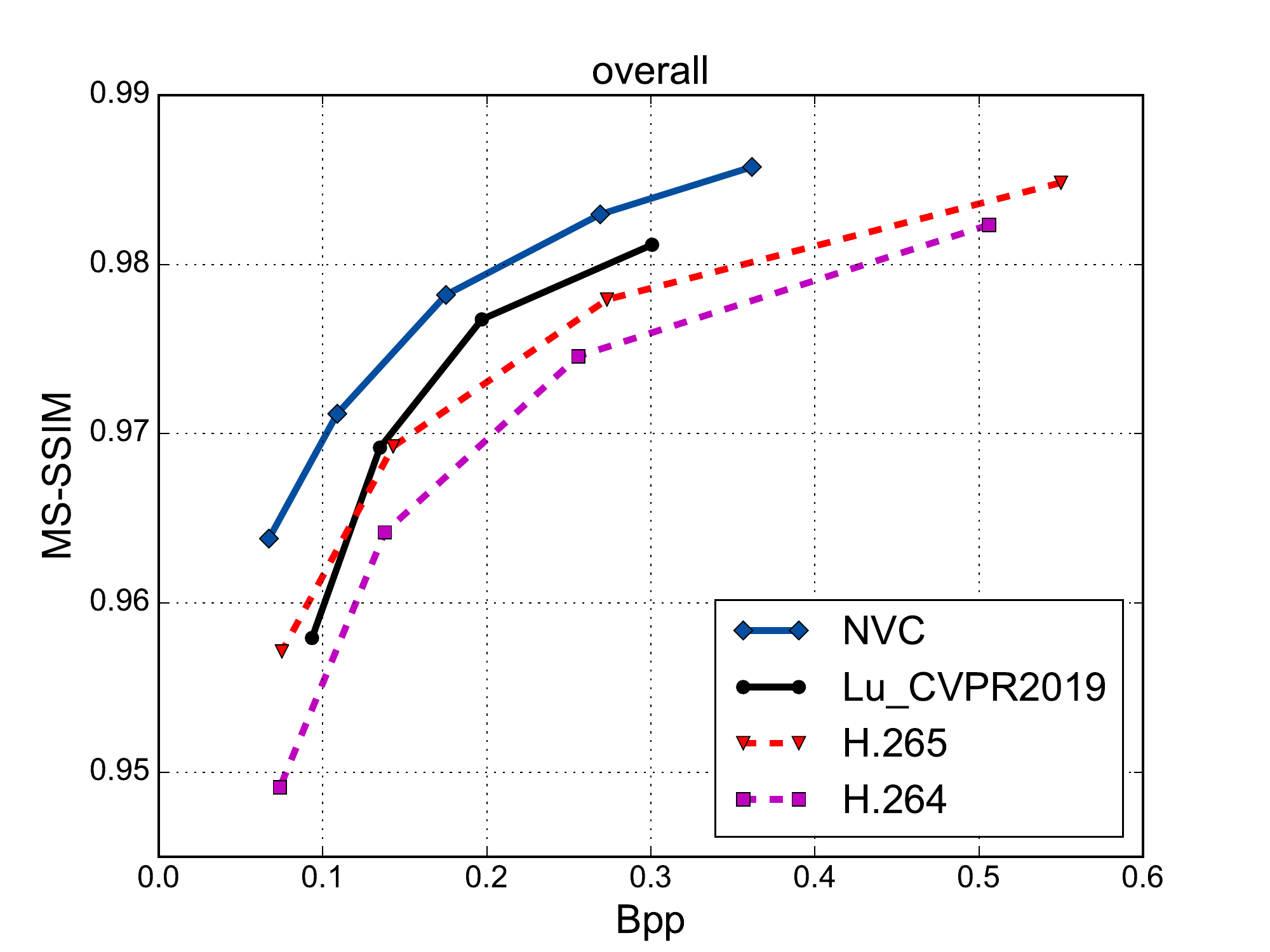}}
\caption{{\bf MS-SSIM vs. Rate Performance.} NVC shows significant gains for all the testing videos. MS-SSIM is usually
more correlated with the perceptual quality than PSNR, especially at low bit rates.
}
\label{rd_curve}
\end{figure*}
\begin{table*}[t]
  \centering
  \caption{BD-Rate Gains of NVC, HEVC and DVC against the H.264/AVC. }
  \begin{tabular}{c|c|c|c|c|c|c|c|c|c|c|c|c}
    \hline
    \multirow{4}{*}{\textbf{Sequences}}&\multicolumn{4}{c|}{\textbf{H.265/HEVC}}&\multicolumn{4}{c|}{\textbf{DVC}}&\multicolumn{4}{c}{\textbf{NVC}}\\
     
    \cline{2-13}
      & \multicolumn{2}{c|}{PSNR} & \multicolumn{2}{c|}{MS-SSIM} &\multicolumn{2}{c|}{PSNR} & \multicolumn{2}{c|}{MS-SSIM} & \multicolumn{2}{c|}{PSNR} & \multicolumn{2}{c}{MS-SSIM} \\
      \cline{2-13}
    & {BDBR} & {BD-(D)} &{BDBR} & {BD-(D)} & {BDBR} & {BD-(D)}&{BDBR} &{BD-(D)} & {BDBR} & {BD-(D)} & {BDBR}& {BD-(D)}\\
    
     \hline
     
    {ClassB} & {-32.03\%} &{0.78} & {-27.67\%} & {0.0046} & {-27.92\%}& {0.72}& {-22.56\%}& {0.0049}& 
\bf{-45.66\%}& 
\bf{1.21}& 
\bf{-54.90\%}& 
\bf{0.0114}\\
    \hline
     {ClassC} & \bf{-20.88\%} & \bf{0.91} & {-19.57\%} & {0.0054} & {-3.53\%}& {0.13}& {-24.89\%}& {0.0081}& {-17.82\%}& 
{0.73}& 
\bf{-43.11\%}& 
\bf{0.0133}\\
    \hline
    {ClassD} & {-12.39\%} &{0.57} & {-9.68\%} & {0.0023} & {-6.20\%}& {0.26}& {-22.44\%}& {0.0067}& 
\bf{-15.53\%}& 
\bf{0.70}& 
\bf{-43.64\%}& 
\bf{0.0123}\\
    \hline
    {ClassE} & {-36.45\%} &{0.99} & {-30.82\%} & {0.0018} & {-35.94\%}& {1.17}& {-29.08\%}& {0.0027}& 
\bf{-49.81\%}& 
\bf{1.70}& 
\bf{-58.63\%}& 
\bf{0.0048}\\
    \hline
    {UVG} & {-48.53\%} &{1.00} & {-37.5\%} & {0.0056} & {-37.74\%}& {1.00}& {-16.46\%}& {0.0032}& 
\bf{-48.91\%}& 
\bf{1.24}& 
\bf{-53.87\%}& 
\bf{0.0100}\\
    \hline
    {Average} & {-30.05\%} &{0.85} & {-25.04\%} & {0.0039} & {-22.26\%}& {0.65}& {-23.08\%}& {0.0051}& \bf{-35.54\%}& \bf{1.11}& \bf{-50.83\%}& \bf{0.0103}\\

    \hline
  \end{tabular}
  \label{tab:BDrate}
\end{table*}
\begin{figure*}[t]
     \centering
     \includegraphics[scale=0.21]{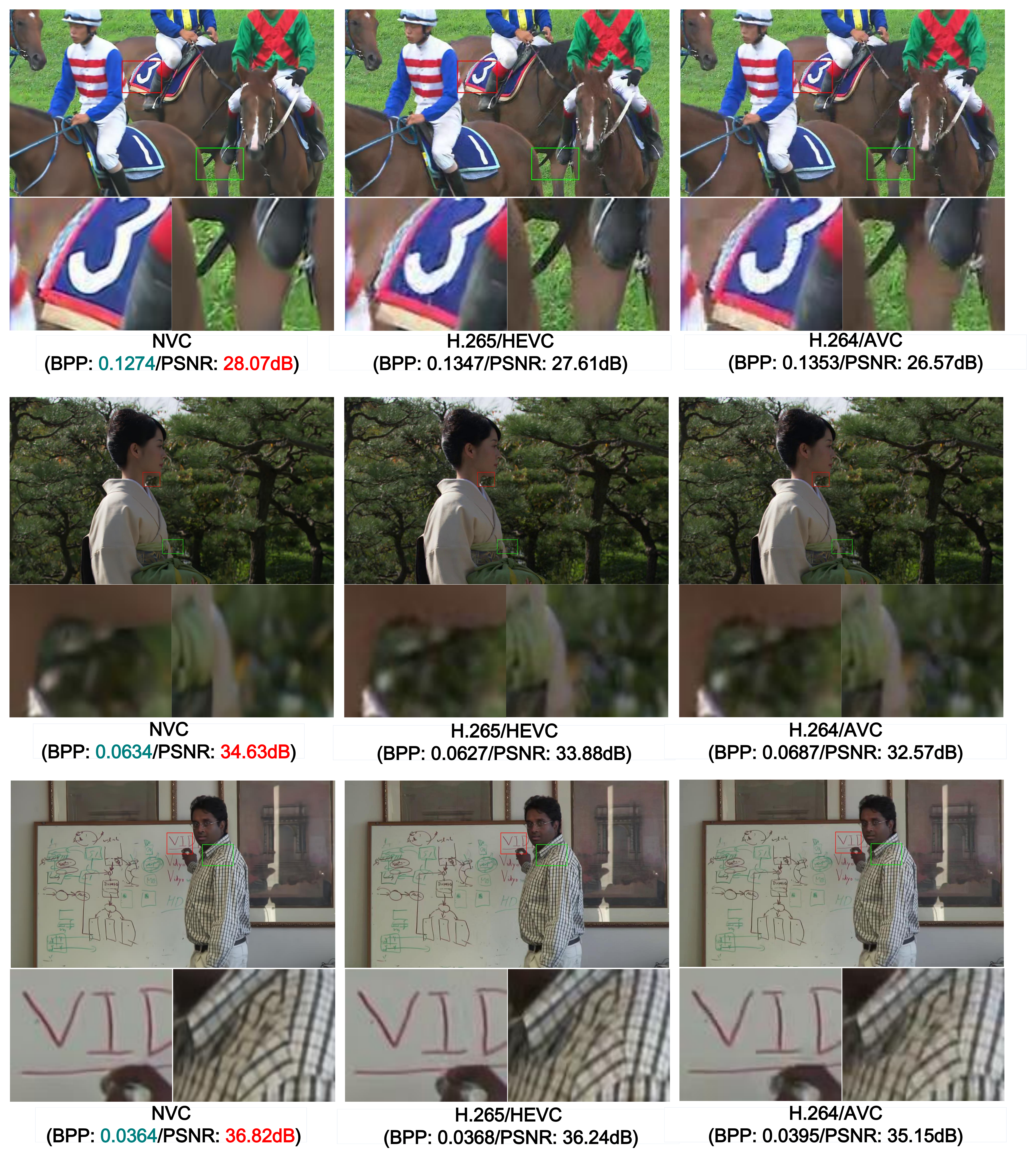}
     \caption{{\bf Visual Comparison.} Reconstructed frames of NVC, HEVC and H.264/AVC. NVC has fewer blocky artifacts and visible noise, etc, and provides better quality at lower bit rate.}
     \label{fig:visual_comparison}
\end{figure*}

{\bf Rate distortion Performance.} We compare the coding perfomance of different methods in  Fig.~\ref{rd_curve_psnr} and~\ref{rd_curve} using respective PSNR and MS-SSIM measures, across HEVC and UVG test sequences. Note that when we report PSNR or MS-SSIM values, the models are trained using PSNR and MS-SSIM, respectively, as the distortion metric.
In Table~\ref{tab:BDrate}, by setting the same anchor using H.264/AVC, our NVC presents 35\% BD-Rate gains when the distortion is measured by PSNR, while HEVC offers 30\% gains. Gains are even larger, if the distortion is measured by the MS-SSIM. In this case, NVC can achieve 50\% improvement, while HEVC is around 25\%. 

Wu et al.~\cite{wu2018vcii} proposed a interpolation based video coding framework and could not get better performance than H.265/HEVC. Our NVC performs significantly better than DVC~\cite{Lu_2019_CVPR}, which has 22.26\% and 23.08\% BD-Rate reductions under PSNR and MS-SSIM metrics, respectively. From Fig.~\ref{rd_curve_psnr} and Fig.~\ref{rd_curve}, DVC~\cite{Lu_2019_CVPR} has mainly improved the coding efficiency against HEVC at high bit rates. However, DVC is not competitive at low bit rate (e.g., having performance even worse than H.264/AVC at some rate regimes.). We have also observed that DVC's performance varies for different test sequences. An improved version of DVC, known as DVC\_pro, has been recently reported, which has shown the state of the art performance~\cite{lu2020end} by using~\cite{minnen2018joint} for intra and residual coding and $\lambda$ tuning. NVC is slightly better
that DVC Pro in both PSNR and MS-SSIM,
at 34.5\% and 45.88\%, respectively.

{\bf Visual Comparison.}
We provide the visual quality comparison with H.264/AVC and H.265/HEVC in Fig.~\ref{fig:visual_comparison}.  Generally,  NVC leads to a higher quality reconstruction at slightly lower bit rate. For example, for the ``RaceHorse'' which has nontranslational motion and complex background, NVC uses 7\% less bits for more than 1.5 dB PSNR improvement compared with H.264/AVC. For other cases, our method also shows robust improvement.
Traditional codec usually suffers from blocky artifacts and motion-induced noise close to object edges. One can clearly observe block partition boundaries with severe pixel discontinuity in reconstructed frames by H.264/AVC. Our results have much less visible noise and artifacts.



\subsection{Ablation Study}
This section examines modular components in NVC to further understand its capacity for application.

{\bf Spatiotemporal Context Modeling.} To evaluate the gain from using temporal priors generated by ConvLSTM for the entropy coding of the motion features, we have also trained a  model when the PA engine in neuro-Motion does not use the temporal priors. Table~\ref{tab:rd_abl} shows that 2\% to 7\% rate increase are incurred if the temporal priors are not used. This reveals that temporal priors help to make probability prediction more accurate, leading to less bit consumption for compressing the motion features.
\begin{table}[t]
  \centering
  \caption{ Efficiency of Temporal Priors for Coding Motion Features. The entries in 2nd and 3rd columns are the BD-rate reduction.}

  \begin{tabular}{|c|c|c|c|}
    \hline
       Sequences & NVC & W/O temporal priors & Bits saving \\
    \hline
     ClassB & -54.90\% & -48.77\% & 6.13\%\\
     \hline
     ClassC & -43.11\% & -39.17\%& 3.94\%\\
    \hline
     ClassD & -43.64\% & -39.23\% & 4.41\% \\
    \hline
     ClassE & -58.63\% & -56.27\%& 2.36\% \\
     \hline
     UVG dataset & -53.87\%& -49.14\%& 4.73\% \\
     \hline
     Average & -50.83\%& -46.52\%& 4.31\%\\
    \hline
  \end{tabular}
  \label{tab:rd_abl}
\end{table}

 Generally, bits consumed by motion information  vary across different content and total target rates. More saving is reported for low bit rates, and for motion intensive content. In these cases, motion bits usually occupy a higher percentage of the total rate. For stationary content, such as HEVC Class E, motion bits contribute less percentage, thus the gain from using temporal priors is less significant. 
 
 \begin{figure*}[t]
     \centering
     \includegraphics[scale=0.43]{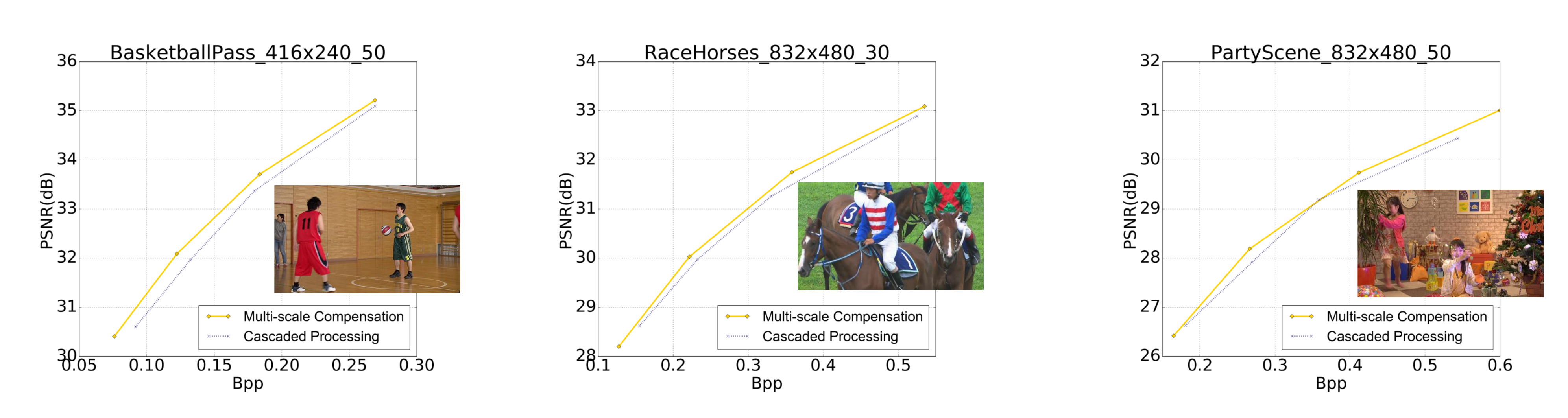}
     \caption{{\bf Comparison with using a Cascaded Denoising Network.} Proposed MS-MCN can achieve better rate-distortion performance compared with using a denoising network following single scale motion compensation.  }
     \label{fig:comp_processing}
\end{figure*}

\begin{figure}[t]
     \centering
     \includegraphics[scale=0.20]{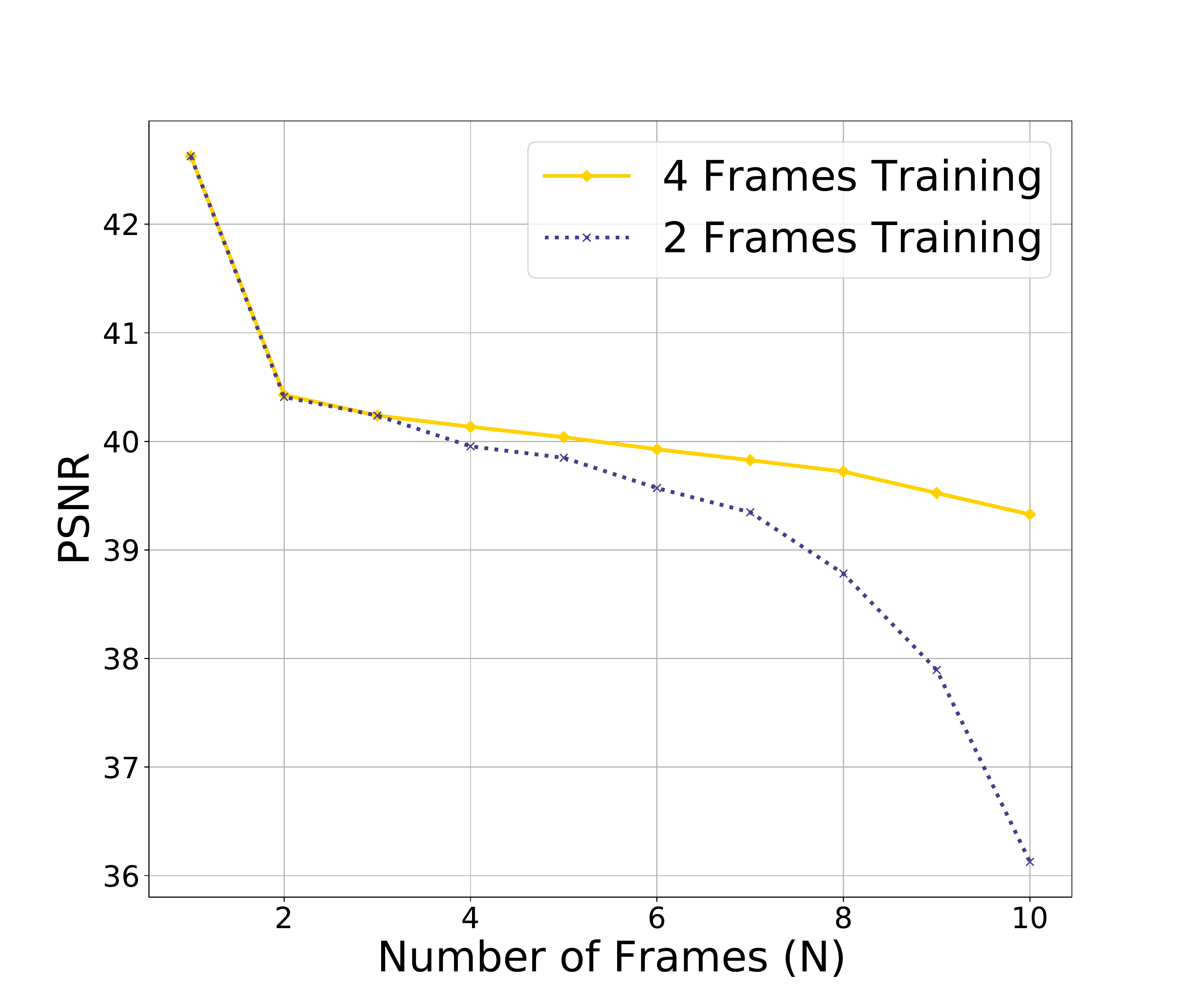}
     \includegraphics[scale=0.20]{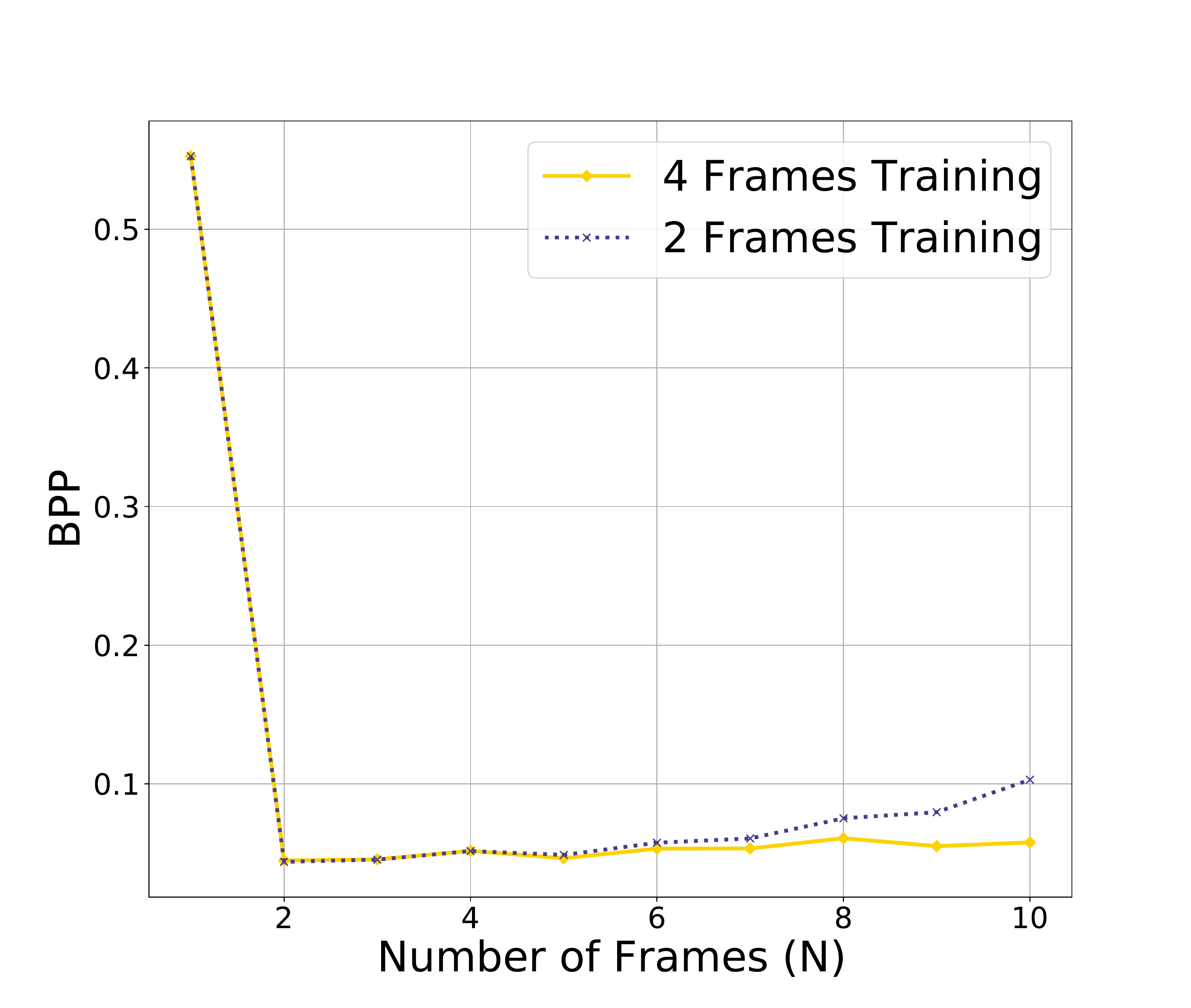}
     \caption{{\bf Benefit from Multi-frame Training.} Two-frame training could only learn temporal variations from intra to inter reconstruction; Multi-frame training is applied to capture both intra-to-inter, and  inter-to-inter variations, yielding improved stability and performance.}
     \label{fig:multi_frame_training}
\end{figure}

{\bf Comparison with Cascaded Denoising Networks.} 
Linear warping often brings artifacts (e.g., ghosting edges), especially when flow estimation is not accurate and occlusion happens.
To remove such artifacts, a denoising network is often added after motion compensated warping as shown in Fig.~\ref{sfig:two_stage} and~\ref{sfig:one_stage}.
Our MS-MCN, instead, rely on multiscale motion fields to generate the predicted frame. 
As a comparison, we also trained a U-net like denoising network on the warped frames based on the single-scale decoded motion field. As shown in Fig.~\ref{fig:comp_processing}, MS-MCN achieves 0.2 to 0.3 dB PSNR gains across a large bit rate ranges over using a cascaded denoising network  for various content for overall performance.

{\bf Temporal Stability with Multi-frame Training.} 
Compression Errors (e.g., lossy compression noise) often propagate from one frame to another due to predictive coding structure. It is critical to limit the temporal quality degradation during training. 
To evaluate the gain from multi-frame training, we compare the performance of the NVC model trained by minimizing reconstruction errors for only one P-frame (single frame training) and the model further refined by minimizing the reconsturction loss for multiple consecutive P-frames. We have varied the number of total frames to use while training, and found that using 4 frames reaches a good compromise. As shown in Fig.~\ref{fig:multi_frame_training}, multi-frame training could well capture temporal quality variations from intra to inter, and from inter to inter, leading to a more generalized model with consistent performance  for all the frames in a GOP with slower quality degradation, and improved stability.


%% file: conc.tex
\section{Conclusion}\label{sec:conc}

We have developed an end-to-end deep neural video coding framework (NVC) that can compactly represent the intra-pixel, inter-motion and inter-residual information, respectively. We have shown that the pyramid decoder in neuro-Motion and the multiscale motion compensation network (MS-MCN) together can significantly improve inter-frame prediction, compared to the more conventional single scale motion compensation. Furthermore, adding temporal context  can lead to more efficient entropy coding of the motion information than  using only spatial context and hyper priors. We further demonstrate that using a multi-frame training loss can effectively mitigate the temporal error propagation.

We evaluate the NVC by PSNR and MS-SSIM respectively, and compare its performance both with standard coding methods including H.264/AVC and H.265/HEVC as well as learnt video coders including Wu et al.~\cite{wu2018vcii}, DVC~\cite{Lu_2019_CVPR} and DVC\_pro~\cite{lu2020end}.
NVC offers consistent and stable gains over existing methods across a variety of contents and bit rates.

The proposed DNN-based model can be further improved by engineering better stacked CNNs instead of current implementation. 
Context-adaptive probability models could be further improved. For example, recent exploration in~\cite{lee2019end} has demonstrated that weighted GMM could further improve the entropy coding efficiency. This can be easily incorporated in our PA engine.
Reference frame selection is another direction for overall efficiency optimization, by which we can embed and aggregate most appropriate information for improving the inter-coding efficiency~\cite{li2012rate}. 

The H.264/AVC, HEVC, and even VVC, are engineering masterpieces for video coding. $\lambda$ adaptation, rate control, etc can be definitely borrowed to improve the NVC. Furthermore, how to make NVC practically applicable is also worth for deep investigation.

 To benefit the research community,  all relevant materials will be made publically available soon at {\url{https://njuvision.github.io/Neural-Video-Coding}.

%% file: ack.tex
\section{Acknowledgement}
Our gratitude is directed to anonymous reviewers for their suggestions and efforts.